# Robust Agent Teams via Socially-Attentive Monitoring

**Gal A. Kaminka**                                                    GALK@ISI.EDU
**Milind Tambe**                                                      TAMBE@ISI.EDU
*Information Sciences Institute and Computer Science Department*
*University of Southern California*
*4676 Admiralty Way*
*Los Angeles, CA 90292, USA*

## Abstract

Agents in dynamic multi-agent environments must monitor their peers to execute individual and group plans. A key open question is how much monitoring of other agents' states is required to be effective: *The Monitoring Selectivity Problem*. We investigate this question in the context of detecting failures in teams of cooperating agents, via *Socially-Attentive Monitoring*, which focuses on monitoring for failures in the social relationships between the agents. We empirically and analytically explore a family of socially-attentive teamwork monitoring algorithms in two dynamic, complex, multi-agent domains, under varying conditions of task distribution and uncertainty. We show that a centralized scheme using a complex algorithm trades correctness for completeness and requires monitoring all teammates. In contrast, a simple distributed teamwork monitoring algorithm results in correct and complete detection of teamwork failures, despite relying on limited, uncertain knowledge, and monitoring only key agents in a team. In addition, we report on the design of a socially-attentive monitoring system and demonstrate its generality in monitoring several coordination relationships, diagnosing detected failures, and both on-line and off-line applications.

## 1. Introduction

Agents in complex, dynamic, multi-agent environments must be able to detect, diagnose, and recover from failures at run-time (Toyama & Hager, 1997). For instance, a robot's grip may be slippery, opponents' behavior may be intentionally difficult to predict, communications may fail, etc. Examples of such environments include virtual environments for training (Johnson & Rickel, 1997; Calder, Smith, Courtemanche, Mar, & Ceranowicz, 1993), high-fidelity distributed simulations (Tambe, Johnson, Jones, Koss, Laird, Rosenbloom, & Schwamb, 1995; Kitano, Tambe, Stone, Veloso, Coradeschi, Osawa, Matsubara, Noda, & Asada, 1997), and multi-agent robotics (Parker, 1993; Balch, 1998). The first key step in this process is execution-monitoring (Doyle, Atkinson, & Doshi, 1986; Ambros-Ingerson & Steel, 1988; Cohen, Amant, & Hart, 1992; Reece & Tate, 1994; Atkins, Durfee, & Shin, 1997; Veloso, Pollack, & Cox, 1998).

Monitoring execution in multi-agent settings requires an agent to monitor its peers, since its own correct execution depends also on the state of its peers (Cohen & Levesque, 1991; Jennings, 1993; Parker, 1993; Jennings, 1995; Grosz & Kraus, 1996; Tambe, 1997). Monitoring peers is of particular importance in teams, since team-members rely on each other and work closely together on related tasks:

- Monitoring allows team-members to coordinate their actions and plans with team-mates, to help teammates and cooperate without interference. For example, drivers of cars in a convoy cannot drive without monitoring other cars in the convoy, so as to not disband the convoy, and to help other drivers if cars break down.

- Monitoring allows team-members to use peers as dynamic information sources, for learning new information. For instance, if a driver in a convoy sees that the other cars in front of it suddenly turn to the left, she can infer the existence of an obstacle or milestone despite not directly seeing it herself.

Previous work has investigated different ways of monitoring in the context of teams of co-operating agents. For example, theoretical work on SharedPlans (Grosz & Kraus, 1999) has distinguished between *passive monitoring*, in which an agent is notified when a proposition changes (e.g., via communications), and *active monitoring*, in which an agent actively seeks to find out when a proposition changes (e.g., via observations and inference of unobservable attributes). Practical implementations have investigated the use of passive monitoring via communications (Jennings, 1995), active monitoring via plan-recognition (Huber & Durfee, 1995), active implicit monitoring via the environment (Fenster, Kraus, & Rosenschein, 1995), and different combinations of these methods (Parker, 1993; Jennings, 1993; Tambe, 1997; Lesh, Rich, & Sidner, 1999). No approach is clearly superior to another: Passive monitoring is generally perceived as being less costly than active monitoring, but also less reliable (Grosz & Kraus, 1999; Huber & Durfee, 1995; Kaminka & Tambe, 1998).

Regardless of the monitoring method, bandwidth and computational limitations prohibit a monitoring agent from monitoring all other agents to full extent, all the time (Jennings, 1995; Durfee, 1995; Grosz & Kraus, 1996). Thus a key open question is how much monitoring of other agents is required to be effective (in teams) (Jennings, 1993; Grosz & Kraus, 1996, 1999). We call this challenging problem the *Monitoring Selectivity Problem*, i.e., the problem of selectivity in observing others and inferring their state (based on the observations) for monitoring. Although it has been raised in the past, only a framework and minimal constraints for answers were provided (Jennings, 1993; Grosz & Kraus, 1996). For instance, the theory of SharedPlans requires agents to verify that their intentions do not conflict with those of teammates (Grosz & Kraus, 1996). However, the methods by which such verification can take place are left for further investigation (Grosz & Kraus, 1996, p. 308). Section 8 provides more details on related work.

This paper begins to address the monitoring selectivity problem in teams, by investigating monitoring requirements for effective failure detection. We focus our investigation on detecting failures in the social relationships that ideally hold between agents in a monitored team. We call such monitoring of social relationships *socially-attentive monitoring*, to differentiate it from other types of monitoring, such as monitoring for failures in the progress of agents towards their goals. Here, the term social relationship is used to denote a relation on attributes of multiple agents' states. Socially-attentive monitoring in the convoy example involves verifying that agents have common destination and heading, that their beliefs in driving as a convoy are mutual, etc. For instance, if the agents are observed to head in different directions, they clearly do not have a common heading. This is different than monitoring whether their chosen (common) heading leads towards their (agreed upon) destination.

Monitoring relationships in a team (socially-attentive monitoring) is a critical task in monitoring team-members. Failures to maintain the team's relationships can often lead to catastrophic failures on the part of the team, lack of cooperative behavior and lack of coordination. Such failures are often the result of individual agent failures, such as failures in an agent's sensors and actuators. Thus socially-attentive monitoring covers a large class of failures, and promotes robust individual operation.

We explore socially-attentive monitoring algorithms for detecting teamwork failures under various conditions of uncertainty. We analytically show that despite the presence of uncertainty about the actual state of monitored agents, a centralized *active* monitoring scheme can guarantee failure detection that is either sound and incomplete, or complete and unsound. However, this requires reasoning about multiple hypotheses as to the actual state of monitored agents, and monitoring all agents in the team. We show that active distributed teamwork monitoring results in both sound and complete detection capabilities, despite using a much simpler algorithm. This distributed algorithm: (a) uses only a single, possibly incorrect hypothesis of the actual state of monitored agents, and (b) involves monitoring only key agents in a team, not necessarily all team-members. Using a transformation on the analytical constructs, we show analogous results for centralized failure-detection in mutual-exclusion coordination relationships.

We also conduct an empirical investigation of socially-attentive monitoring in teams. We present an implemented general socially-attentive monitoring framework in which the expected ideal social relationships that are to be maintained by the agents are compared to the actual social relationships. Discrepancies are detected as possible failures and diagnosed. We apply this framework to two different complex, dynamic, multi-agent domains, in service of monitoring various social relationships, both on-line and off-line. Both of these domains involve multiple interacting agents in collaborative and adversarial settings, with uncertainties in both perception and action. In one domain, we provide empirical results for active monitoring which confirm our analytical results. In another domain we show how off-line socially-attentive monitoring can provide quantitative teamwork quality feedback to a designer. We also provide initial diagnosis procedures for detected failures.

Our focus in these explorations is on practical algorithms that have guarantees on performance in real-world applications. The algorithms we present seek to complement the use of passive communications-based monitoring (which is unreliable in many domains) and explore the use of unintrusive key-hole plan-recognition as an alternative. However, we do not rule out the use of communications—we simply seek to provide techniques that can work even when communications fail. Our analytical guarantees of failure-detection soundness and completeness hold whether monitoring is done through communications or plan-recognition.

This paper is organized as follows: Section 2 presents motivating examples and background. Section 3 presents the socially-attentive monitoring framework. Section 4 explores monitoring selectivity in centralized teamwork monitoring. Section 5 explores monitoring selectivity in distributed teamwork monitoring. Section 6 demonstrates the generality of our framework by applying it in an off-line configuration. Section 7 presents investigations of additional relationship models. Section 8 presents related work, and Section 9 concludes. The two appendices contain the proofs for theorems presented (Appendix A), and pseudo-code for the socially-attentive monitoring algorithms (Appendix B).

## 2. Motivation and Background

The monitoring selectivity problem this paper addresses—how much monitoring is required for failure-detection in teams—rose out of growing frustration with the significant software maintenance efforts in two of our application domains. In the ModSAF domain, a high-fidelity battlefield virtual environment (Calder et al., 1993), we have been involved in the development of synthetic helicopter pilots (Tambe et al., 1995). In the RoboCup soccer simulation domain (Kitano et al., 1997) we have been involved in developing synthetic soccer players (Marsella, Adibi, Al-Onaizan, Kaminka, Muslea, Tallis, & Tambe, 1999). The environments in both domains are dynamic and complex, and have many uncertainties: the behavior of other agents (some of it adversarial, some cooperative), unreliable communications and sensors, actions which may not execute as intended, etc. Agents in these environments are therefore presented with countless opportunities for failure despite the designers' best efforts.

Some examples may serve to illustrate. The following two examples are actual failures that occurred in the ModSAF domain. We will use these two to illustrate and explore socially-attentive monitoring throughout this paper:

**Example 1.** Here, a team of three helicopter pilot agents were to fly to a specified way-point (a given position), where one of the team-members, the *scout*, was to fly forward towards the enemy, while its teammates (*attackers*) land and wait for its signal. All of the agents monitored for the way-point. However, due to an unexpected sensor failure, one of the attackers failed to sense the way-point. So while the other attacker correctly landed, the failing attacker continued to fly forward with the scout (see Figure 1 for a screen shot illustrating this failure).

**Example 2.** In a different run, after all three agents reached the way-point and detected it, the scout has gone forward and identified the enemy. It then sent a message to the waiting attackers to join it and attack the enemy. One of the attackers did not receive the message, and so it remained behind indefinitely while the scout and the other attacker continued the mission alone.

We have collected dozens of similar reports in both the ModSAF and RoboCup domain. In general, such failures are difficult to anticipate in design time, due to the huge number of possible states. The agents therefore easily find themselves in novel states which have not been foreseen by the developer, and the monitoring conditions and communications in place proved insufficient: In none of the failure cases reported did the agents involved detect, let alone correct, their erroneous behavior. Each agent believed the other agents to be acting in coordination with it, since no communication was received from the other agents to indicate otherwise. However, the agents were violating the collaboration relationships between them, as the agents came to disagree on what plan is being executed—a collaboration relationship failure had occurred. Preliminary empirical results show that upwards of 30% of failures reported involved relationship violations (relationship failures).

Human observers, however, were typically quick to notice these failures, because of the clear social misbehavior of the agents in these cases. They were able to infer that a failure has occurred despite not knowing what exactly happened. For instance, seeing an attacker continuing to fly ahead despite its teammates' switching to a different plan (which the human observers inferred from the fact that one of the teammates, the other attacker, has landed) is sufficient for an observer to detect that something has gone amiss—without knowing what the different plan was.

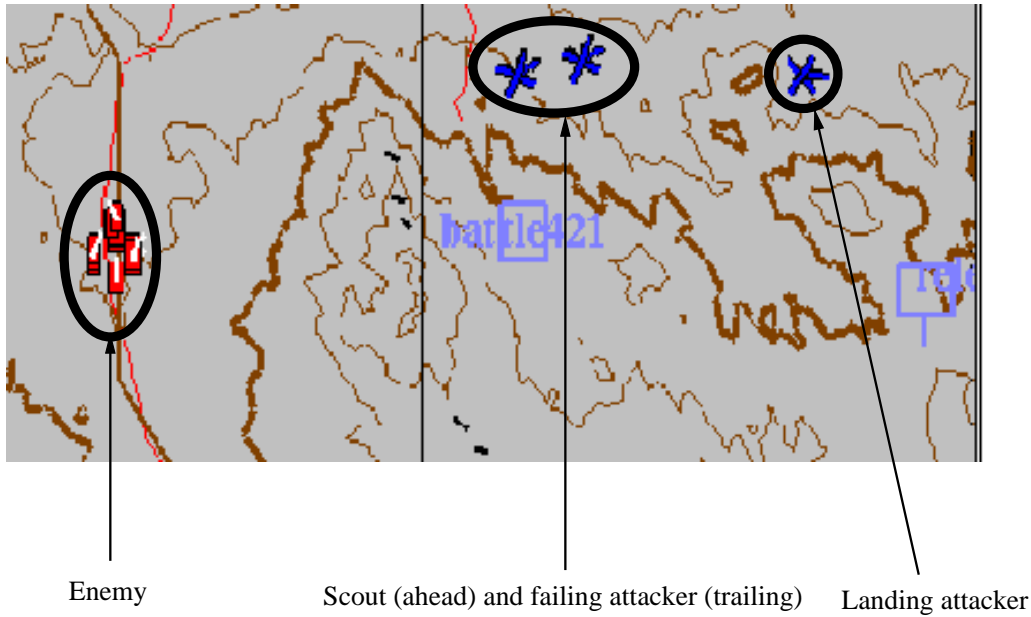

Figure 1: A plan-view display (the ModSAF domain) illustrating the failure in Example 1. The thick wavy lines are contour lines.

Our analysis showed that the agents were not monitoring each other sufficiently. However, a naive solution of continuous communications between the agents was clearly impractical since: (i) the agents are operating in a hostile environment; (ii) the communications overheads would have been prohibitive; and (iii) in fact, it was the communications equipment itself that broke down in some cases. We therefore sought practical ways to achieve quick detection of failure, based on the limited ambiguous knowledge that was available to a monitoring agent.

## 3. Socially-Attentive Monitoring

We begin with an overview of the general structure of a socially-attentive monitoring system, shown in Figure 2. It consists of: (1) a social relationship knowledge-base containing models of the relationships that should hold among the monitored agents, enabling generation of *expected ideal behavior* in terms of relationships (Section 3.1); (2) an agent and team modeling component, responsible for collecting and representing knowledge about the monitored agents' *actual behavior* (Section 3.2); (3) a relationship failure-detection component that monitors for violations of relationships among monitored agents by contrasting the expected and actual behavior (Section 3.3); and (4) a relationship diagnosis component that verifies the failures, and provides an explanation for them (Section 3.4). The resulting explanation (diagnosis) is then used for recovery, e.g., by a negotiations system (Kraus, Sycara, & Evenchik, 1998), or a general (re)planner (Ambros-Ingerson & Steel, 1988).

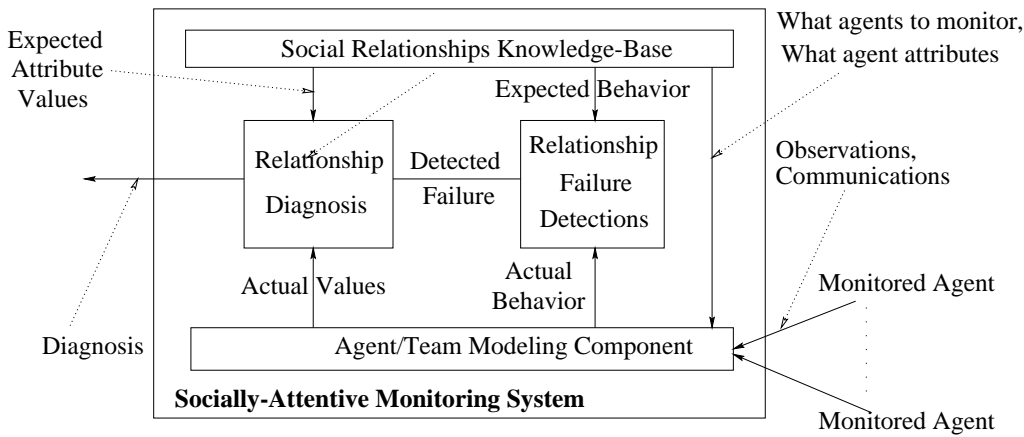

Figure 2: The general structure of a socially-attentive monitoring system.

## 3.1 A Knowledge-Base of Relationship Models

We take a relationship among agents to be a relation on their state attributes. A relationship model thus specifies how different attributes of an agent's state are related to those of other agents in a multi-agent system. These attributes can include the beliefs held by the agents, their goals, plans, actions, etc. For example, many teamwork relationship models require that team-members have mutual belief in a joint goal (Cohen & Levesque, 1991; Jennings, 1995). A spatial formation relationship (Parker, 1993; Balch, 1998) specifies relative distances, and velocities that are to be maintained by a group of agents (in our domain, helicopter pilots). Coordination relationships may specify temporal relationships that are to hold among the actions of two agents, e.g., business contractors (Malone & Crowston, 1991). All such relationships are *social*—they explicitly specify how multiple agents are to act and what they are to believe if they are to maintain the relationships between them.

The relationship knowledge-base contains models of the relationships that are supposed to hold in the system, and specifies the agents that are participating in the relationships. The knowledge-base guides the agent-modeling component in selecting agents to be monitored, and what attributes of their state need be represented (for detection and diagnosis). It is used by the failure detection component to generate expectations which are contrasted with actual relationships maintained by the agents. And it provides the diagnosis component with detailed information about how agents' states' attributes are related, to drive the diagnosis process. Our implementation of socially-attentive monitoring in teams uses four types of relationships: *formations*, *role-similarity*, *mutual exclusion*, and *teamwork*.

For teamwork monitoring we use the STEAM (Tambe, 1997) general domain-independent model of teamwork, which is based on Cohen and Levesque's Joint Intentions Framework (Levesque, Cohen, & Nunes, 1990; Cohen & Levesque, 1991) and Grosz, Sidner, and Kraus's SharedPlans (Grosz & Sidner, 1990; Grosz & Kraus, 1996, 1999). However, other teamwork models may be used instead of STEAM. Although STEAM is used by our pilot and soccer agents to generate collaborative behavior, it is reused here independently in service of monitoring, i.e., monitored agents are assumed to be a team, and STEAM is used in monitoring their teamwork. STEAM and other teamwork models (e.g., Cohen & Levesque, 1991; Jennings, 1995; Rich & Sidner, 1997) require mutual belief by team members in their joint goals and plans. This characteristic is used to monitor teamwork in our system. The other relationship models are used only in a secondary monitoring role. They will be discussed in greater length in Section 7.

## 3.2 Knowledge of Monitored Agents and Team

The agent modeling component is responsible for acquiring and maintaining knowledge about monitored agents. This knowledge is used to construct the *actual* relations that exist between agents' states' attributes, which are compared to the *ideal expected* relations. In this section, we describe the plan-recognition capabilities of the agent-modeling component in our implementation and experiments, i.e., the extent of the knowledge that could be maintained about monitored agents' plans if necessary. Later sections show that in fact limited, possibly inaccurate, knowledge is sufficient for effective failure detection. Thus implementations may use optimized agent-modeling algorithms rather than these full capabilities. Section 3.4 will discuss additional agent-modeling capabilities, necessary for diagnosis.

### 3.2.1 Representation

For monitoring teamwork relationships, we have found that representing agents in terms of their selected hierarchical reactive plans enables quick monitoring of their state, and also facilitates further inference of the monitored agents' beliefs, goals, and unobservable actions, since they capture the agents' decision processes.

In this representation, reactive plans (Firby, 1987; Newell, 1990) form a single decomposition hierarchy (a tree) that represents the alternative controlling processes of each agent. Each reactive plan in the hierarchy (hereafter referred to simply as a plan) has selection conditions (also referred to as preconditions) for when it is applicable, and termination conditions which are used to terminate or suspend plans. At each given moment, the agent is executing a single path (root to a leaf) through the hierarchy. This path is composed of plans at different levels.

Figure 3 presents a small portion of such a hierarchy, created for the ModSAF domain. In the case of Example 1, prior to the way-point, *each of the agents* was executing the path beginning with `execute-mission` as highest-level plan, through `fly-flight-plan`, `fly-route`, `traveling` and `low-level`. Upon reaching the way-point, they were all supposed to switch from `fly-flight-plan` and its descendents to `wait-at-point`. The attackers would then select `just-wait` as a child of `wait-at-point`, while the scout would select `scout-forward` and its descendents. Of course, the failing attacker did not detect the way-point and so the termination conditions for `fly-flight-plan` and the selection conditions for `wait-at-point` were not satisfied and the failing attacker continued to execute `fly-flight-plan` and its descendents.

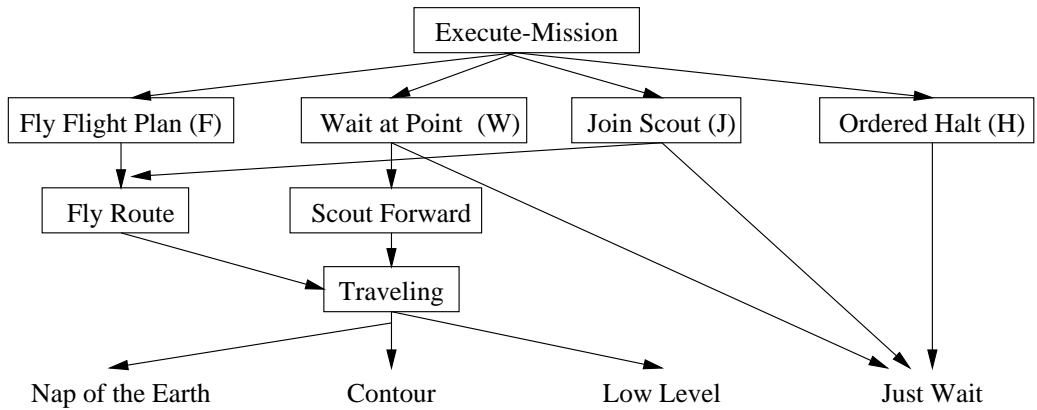

Figure 3: Portion of Hierarchical Reactive Plan Library for ModSAF Domain (Team plans are boxed. These are explained in Section 3.3).

### 3.2.2 ACQUISITION

From a practical perspective, while the agents may cooperatively report to the monitoring agent on their own state using communications, it requires communication channels to be sufficiently fast, reliable and secure. This is unfortunately not possible in many realistic domains, as our examples demonstrate (Section 2).

Alternatively, a monitor may use plan-recognition to infer the agents' unobservable state from their observable behavior. This approach is unintrusive and robust in face of communication failures. Of course, the monitor may still benefit from focused communications with the other agents, but would not be critically dependent on them.

To enable plan-recognition using reactive plans (our chosen representation), we have employed a reactive plan-recognition algorithm called RESL (REal-time Situated Least-commitments). The key capability required is to allow explicit maintenance of hierarchical plan hypotheses matching each agent's observed behavior, while pruning of hypotheses which are deemed incorrect or useless for monitoring purposes. RESL works by expanding the entire plan-library hierarchy for each modeled agent, and tagging all paths matching the observed behavior of the agent being modeled (see Appendix B for pseudo-code for the algorithm). Heuristics and external knowledge may be used to eliminate paths (hypotheses) which are deemed inappropriate—indeed such heuristics will be explored shortly. RESL's basic approach is very similar to previous work in reactive plan recognition (Rao, 1994) and team-tracking (Tambe, 1996), which have been used successfully in the ModSAF domain, and share many of RESL's properties. However, RESL adds belief-inference capabilities which are used in the diagnosis process, discussed below (Section 3.4).

Figure 4 gives a simplified presentation of the plan hierarchies for a variation of Example 1, in which all the agents correctly detected the way-point, i.e., no failure has occurred (note that some plans at intermediate levels have been abstracted out in the figure). The scout (Figure 4a) and the two attackers (Figures 4b, 4c) switched from the `fly-flight-plan` plan (denoted by F) to the `wait-at-point` plan (denoted by W). An outside observer using RESL infers explanations for each agent's behavior by observing the agents. The scout continues to fly ahead, its speed and altitude matching `low-level`, one of the possible flight-methods under both the `fly-flight-plan` (F) and `wait-at-point` (W) plans. Thus they are both tagged as possible hypotheses for the scout's executing plan hierarchy. Similarly, as the attackers land, RESL recognizes that they are executing the `just-wait` plan. However, this plan can be used in service of either the W or the `ordered-halt` (H) plan—a plan in which the helicopters are ordered by their headquarters to land immediately. Thus both H and W are tagged as explanations for each of the attackers' states (at the second level of the hierarchies). For all agents, RESL identifies the plan `execute-mission` as the top-level plan. In this illustration, the actual executing paths of the agents are marked with filled arrows. Other *individual modeling* hypotheses that match the observed behavior are marked using dashed arrows. An outside observer, of course, has no way of knowing which of the possible hypotheses is correct.

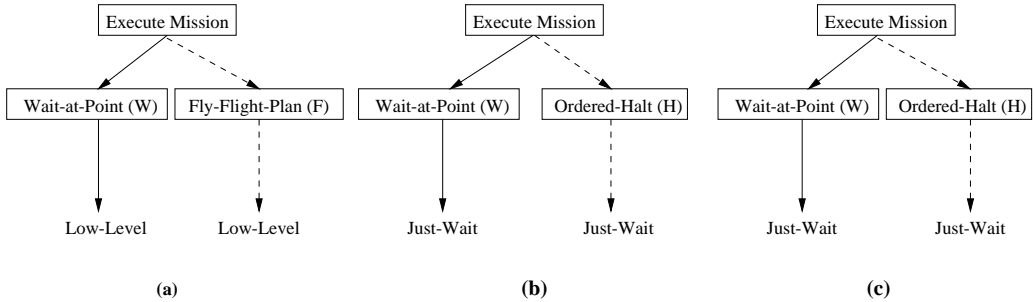

Figure 4: Scout (a) and Attackers' (b, c) actual and recognized *abbreviated* reactive plan hierarchies.

Once individual modeling hypotheses are acquired for each individual agent (using plan-recognition in our implementation, but potentially also by communications), the monitoring agent must combine them to create team-modeling hypotheses as to the state of the team as a whole. The monitoring agent selects a single individual modeling hypothesis for each individual agent and combines them into a single team-modeling hypothesis. Several such team-modeling hypotheses are possible given multiple hypotheses for individual agents. For instance, in Figure 4, while all team-hypotheses will have `execution-mission` as the top-level plan, there are eight different team-hypotheses which can be differentiated by their second-level plan: (W,W,W), (W,W,H), (W,H,W), (W,H,H), (F,W,W), (F,W,H), (F,H,W), (F,H,H). If the observer is a member of the team, it knows what it is executing itself, but would still have multiple-hypotheses about its teammates' states. For instance, if the attacker in Figure 4b is monitoring its teammates, its hypotheses at the second level would be (W,W,W), (W,W,H), (F,W,W), (F,W,H).

To avoid explicitly representing a combinatorial number of hypotheses, RESL explicitly maintains all candidate hypotheses for each agent individually, but not all combinations of individual models as team hypotheses. Instead, these combinations are implicitly represented. Thus the number of hypotheses explicitly maintained grows linearly in the number of agents.

113

### 3.3 Relationship Violation Detection

The failure-detection component detects violations of the social relationships that should hold among agents. This is done by comparing the ideal expected relationships to their actual maintenance by the agents. For teamwork specifically, the relationship model requires team-members to always agree on which *team* plan is jointly executed by the team, similarly to Joint Responsibility (Jennings, 1995), and SharedPlans (Grosz & Kraus, 1996). If this requirement fails in actuality (i.e., the agents are executing different team plans) then a teamwork failure has occurred.

The basic teamwork failure detection algorithm is as follows. The monitored agents' plan-hierarchies are processed in a top-down manner. The detection component uses the teamwork model to tag specific plans as team plans, explicitly representing joint activity by the team (these plans are boxed in Figures 3, 5 and 4). The team-plans in equal depths of the hierarchies are used to create team-modeling hypotheses. For each hypothesis, the plans of different agents are compared to detect disagreements. Any difference found is an indication of failure. If no differences are found, or if the comparison reaches individual plans (non-team, therefore non-boxed in the figures) no failure is detected. Individual plans, which may be chosen by an agent individually in service of team plans are not boxed in these figures, and are handled using other relationships as discussed in Section 7

For instance, suppose the failing attacker from Example 1 is monitoring the other attacker. Figure 5 shows its view of its own hierarchical plan on the left. The path on the right represents the state of the other attacker (who has landed). This state has been inferred in this example from observations made by the monitoring attacker (here, we are assuming that the plan-recognition process has resulted in one correct hypothesis for each agent. We will discuss more realistic settings below). In Figure 5, the difference that would be detected is marked by the arrow between the two plans at the second level from the top. While the failing attacker is executing the `fly-flight-plan` team-plan (on the left), the other attacker is executing the `wait-at-point` team-plan (on the right). The disagreement on which team-plan is to be executed is a failure of teamwork.

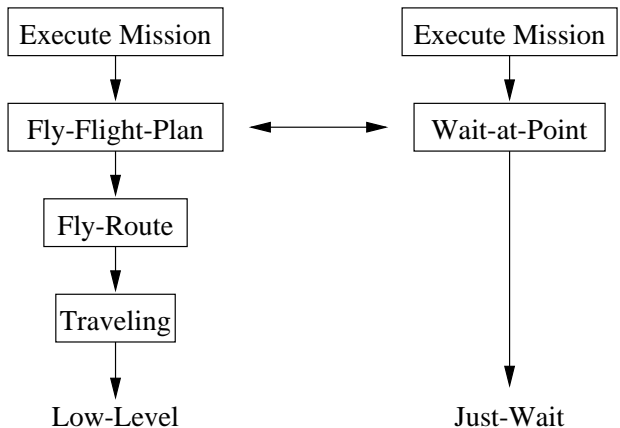

Figure 5: Comparing two hierarchical plans. The top-most difference is at level 2.

Detecting disagreements is difficult with multiple team-modeling hypotheses, since they may imply contradictory results with respect to failure detection: Some hypotheses may imply that a failure had occurred in the team, while others may not. Unfortunately, this is to be expected in realistic applications. For instance, Figure 4 (Section 3.2) shows several hypotheses that are possible based on the same observations. However, one of the hypotheses, (W,W,W), implies no failure has occurred—all the agents are in agreement on which team-plan is executing—while another hypothesis, (F,W,H), implies failures have occurred.

To limit reasoning to only a small number of team hypotheses, while not restricting failure-detection capabilities, we use a disambiguation heuristic that ranks team-modeling hypotheses by the level of *coherence* they represent. This heuristic is provided as an initial solution. Later sections will examine additional heuristics.

**Definition 1.** The *coherence* level of a multi-agent modeling hypothesis is defined as the ratio of the number of agents modeled to the number of plans contained in the hypothesis.

This definition results in a partial ordering of the hypotheses set, from the least coherent hypothesis (one that assigns each agent a different plan than its team-mates), to the most coherent hypothesis (that assigns the same plan to all team members). For instance, the hypothesis (F,W,H) would have the lowest level of coherence, 1, since it implies complete breakdown of teamwork—every agent is executing a different plan. The hypothesis (W,W,W) would have a coherence level of 3, the highest level of coherence for the group of three agents, since they are all assigned the same plan. Ranked between them would be the hypothesis (W,W,H), with a single teamwork failure (disagreement on W and H) and a coherence level of 3/2.

The detection component selects a single maximally-coherent team-modeling hypothesis (ties broken randomly). The intuition for using coherence is that failures to agree occur despite the agents' attempts at teamwork. Thus we expect more agreements than disagreements in the team. The coherence level of a team-hypothesis is inversely related to the number of teamwork failures implied by the hypothesis. Selecting a maximally-coherent hypothesis therefore corresponds to the minimum-number-of-failures heuristic commonly used in diagnosis (Hamscher, Console, & de Kleer, 1992).

For the case depicted in Figure 4, the complete detection process may be conceptualized as follows[1]. Suppose that one of the attackers, whose hierarchy is described in Figure 4b, is monitoring the team. First, it collects the plan hypotheses at the top of the hierarchy for each agent (including itself). In this case, they are {execute-mission}, {execute-mission}, {execute-mission}. Only one team-modeling hypothesis can be built from these: (execute-mission, execute-mission, execute-mission). Since this hypothesis shows no disagreement occurs at this level, the process continues to the second level. Here, the hypotheses for the first agent on the left are {F,W}, for the monitoring second agent (since it knows its own state) there is only one possibility {W}, and for the third agent {W,H}. As we saw above, the maximally team-coherent hypothesis is (W,W,W) which is selected. Since it does not indicate failure, the process continues to the third level. Here the agents are executing individual plans, and so the comparison process stops. Algorithm 2 in Appendix B provides greater details about this process.

---

1. Other implementations may make use of optimized algorithms in which the heuristics are integrated into the agent-modeling algorithm.

When sub-teams are introduced, a difference between team-plans may be explained by the agents in question being a part of different sub-teams. Sub-team members still have to agree between themselves on the joint sub-team plans, but these may differ from one sub-team to the next. For now, let us assume that the teams under consideration are *simple teams*, as defined in Definition 2. We make this definition in service of later analytical results in which it will appear as a condition. We return to the issue of sub-teams in Section 7.1.

**Definition 2.** We say that a team $T$ is *simple*, if its plan-hierarchy involves no different team plans which are to be executed by different sub-teams.

Intuitively, the idea is that in a simple team, all members of the team jointly execute each of the team plans in the hierarchy. This definition is somewhat similar to the definition of a *ground team* in (Kinny, Ljungberg, Rao, Sonenberg, Tidhar, & Werner, 1992), but it does not allow sub-team members of a team to have a joint plan which is different than that of other members.

### 3.4 Relationship Diagnosis

The diagnosis component constructs an explanation for the detected failure, identifying the failure state and facilitating recovery. The diagnosis is given in terms of a set of agent belief differences (inconsistencies) that explains the failure to maintain the relationship. The starting point for this process is the detected failure (e.g., the difference in team-plans). The diagnosis process then compares the beliefs of the agents involved to produce a set of inconsistent beliefs that explain the failure.

Two problems exist in practical applications of this procedure. First, the monitoring agent is not likely to have access to all of the beliefs held by the monitored agents, since it is not feasible in practice to communicate all the agents' beliefs to each other. Second, each agent in a real-world domain may have many beliefs, and many of them will vary among the agents, though most of them will be irrelevant to the diagnosis. Thus relevant knowledge may be simply not be accessible, or may be hidden in mountains of irrelevant facts.

To gain knowledge of the beliefs of monitored agents without relying on communications, the diagnosis process uses a process of belief ascription. The agent-modeling component (using RESL in our implementation) maintains knowledge about the selection and termination conditions of recognized plans (hypotheses). For each recognized plan hypothesis, the modeling component infers that any termination conditions for the plan are believed to be false by the monitored agent (since it has not terminated the plan). We have also found it useful to use an additional heuristic, and infer that the selection conditions (preconditions) for any plan which *has just begun execution* are true. The idea is that when a plan is selected for execution, its preconditions are likely to hold, at least for a short period of time. This heuristic involves an explicit assumption on the part of our system that the new plan is recognized as soon as it begins execution. Designers in other domains will need to verify that this assumption holds.

For each agent $i$, the inferred termination and selection conditions make up a set of beliefs $B_i$ for the agent. For instance, suppose an agent is hypothesized to have just switched from executing `fly-flight-plan` to `wait-at-point`. RESL infers that the agent believes that the way-point was just detected (a selection condition for `wait-at-point`). In addition,

RESL infers that the agent believes that an enemy was not seen, and that no order was received from base to halt the mission (negated termination conditions of `wait-at-point`).

To determine the facts that are relevant to the failure, the diagnosis component uses the teamwork model. The teamwork model dictates which beliefs the agents hold must be mutually believed by all the agents in the team. Any difference that is detected in those beliefs is a certain failure, as the team members do not agree on issues on which agreement is mandatory to participation in the team. The teamwork model thus specifies that the beliefs contained in the $B_i$ sets should be mutual, and should therefore be consistent:

$$\bigcup_i B_i \not\vdash \perp$$

If an inconsistency is detected, the diagnosis procedure looks for contradictions (disagreements) that would cause the difference in team-plan selection. A difference in beliefs serves as the diagnosis, allowing the monitoring agent to initiate a process of recovery, e.g., by negotiating about the conflicting beliefs (Kraus et al., 1998).

For example, as shown in Section 3.3, the two attackers in Example 1 (Section 2) differ in their choice of a team-plan: One attacker is continuing execution of the `fly-flight-plan` plan, in which the helicopters fly in formation. The other attacker has detected the way-point, terminated `fly-flight-plan` and has switched to `wait-at-point`, landing immediately (Figure 5). When the failing attacker monitors its team-mate, it detects a difference in the team-plans (Section 3.3), and the detected difference is passed to diagnosis. The failing attacker makes the following inferences:

1. `Fly-flight-plan` has three termination conditions: (a) seeing the enemy, (b) detecting the way-point, or (c) receiving an order to halt. The failing attacker (left hierarchy in Figure 5) knows its own belief that none of these conditions hold, and thus

$$B_1 = \{\neg WayPoint, \neg Enemy, \neg HaltOrder\}$$

2. `Wait-at-point` has one selection condition: the way-point has been detected. Its termination condition is that the scout has sent a message to join it, having identified the enemy's position. The diagnosis component in this case therefore infers that for the other attacker (right hierarchy in Figure 5)

$$B_2 = \{WayPoint, \neg ScoutMessageReceived\}$$

Then,

$$B_1 \cup B_2 = \{\neg WayPoint, WayPoint, \neg Enemy, \neg ScoutMessageReceived, \neg HaltOrder\}$$

which is inconsistent. The inconsistency (disagreement between the attackers) is $\{\neg WayPoint, WayPoint\}$, i.e., contradictory beliefs about $Waypoint$. Thus now the failing attacker knows that its team-mate has seen the way-point. It can choose to quietly adapt this belief, thereby terminating its own `fly-flight-plan` and selecting `wait-at-point`, or it may choose other recovery actions, such as negotiating with the other attacker on whether the way-point has been reached.

We have found these diagnosis procedures to be useful in many of the failures detected by socially-attentive monitoring (see Section 4 for evaluation and discussion). However, since this paper focuses on the monitoring selectivity problem in *detection*, we leave further investigation of the diagnosis procedures to future work.

## 4. Monitoring Selectivity in Centralized Teamwork Monitoring

Using the socially-attentive framework of Section 3 we systematically examine all failure permutations of Examples 1 and 2 (Section 2) under a centralized teamwork monitoring configuration, where a single team-member is monitoring the team. We vary the agents failing (attacker, attacker and scout, etc.) and the role of the monitoring agent (attacker or scout). We report on the empirical results of detecting and diagnosing failures in all cases. Using these empirical results as a guide, we explore centralized teamwork monitoring analytically. We show that even under monitoring uncertainty, centralized teamwork monitoring can provide either sound or complete detection results (but not both).

As a starting point for our exploration, the monitoring agent uses a single maximally-coherent team-modeling hypothesis as discussed in Section 3.3. We begin with Example 2. The normal order of execution is `wait-at-point` (`W`), followed by `join-scout` (`J`). During the execution of `W`, the two attackers land and wait for the scout to visually identify the enemy's position. Upon identification, the scout sends them a message to join it, which triggers the selection of the `J` plan, and the termination of the `W` plan. When executing `J`, the scout hovers at low altitude, waiting for the attackers to join it. Any failures here are on the part of the attackers (they cannot receive the message) or on the part of the scout (it cannot send it). These failures arise, for instance, if the radio is broken or team-members are out of range. When an agent fails, it continues to execute `W` instead of switching to `J`.

Table 1 summarizes the permutations of Example 2. The permutation number appear in the first column. The next three columns show the actual plans selected by the three agents A1, A2 and A3 in each permutation. The second-to-last column shows whether a relationship failure has occurred in each case, i.e., whether disagreement exists between the agents. Finally, the last column details the physical conditions in each case. There are five possible failure permutations: In case 1, none of the agents failed. In cases 2 and 3 one attacker failed. In case 4 the scout failed to send a message or both attackers failed to receive it. In case 5 the scout does not identify the enemy's position (so no message is sent, and all three agents continue to execute the `W` plan). Other permutations are not possible, since no attacker can switch to the `J` plan without the scout.

For instance, case 2 in Table 1 corresponds to Example 2. The scout (A3) has detected the enemy, switched to plan `J`, and sent a message to the attackers to join it. One attacker (A2) received the message, switched to plan `J`, and began flying towards the scout. However, the remaining attacker (A1) failed to receive the message, and so it maintains its position, continuing to execute `W` and failing to switch to `J`. Since the agents are no longer in agreement on which team plan should be jointly executed, a teamwork failure has occurred. Condition monitors were used in the original failure case to monitor for the scout's message. However failures in communications resulted in these monitoring conditions to be rendered useless.

| Case | Actual Executing Plans | | | Relationship Failure | Physical |
|---|---|---|---|---|---|
| # | Attacker A1 | Attacker A2 | Scout A3 | Occurred? | Failure |
| 1 | J | J | J | - | - |
| 2 | W | J | J | + | A1 fails to receive |
| 3 | J | W | J | + | A2 fails to receive |
| 4 | W | W | J | + | A3's message lost |
| 5 | W | W | W | - | Enemy not identified |

Table 1: All possible failure permutations of the broken radio-link scenario (Example 2).

One key issue is raised by case 5 in Table 1. Here, due to the scout's inability to identify the enemy's position (perhaps due to failure on the scout's part, perhaps because the enemy is simply not there), the three helicopter pilots remain in agreement that the enemy has not been identified. Here, even though clearly the pilots are failing to make progress towards the task goals (the scout continues to search for the enemy indefinitely), no relationship failure is taking place, since the agents are maintaining the teamwork relationship while failing to make progress. This clearly demonstrates that not all failures are necessarily relationship failures.

Table 2 presents the results of the scout monitoring its teammates in Example 2, using a maximally team-coherent hypothesis as the basis for detection. The first column again shows the case number, for reference into Table 1. The next three columns show the scout's (A3's) hypothesis about what plan each agent is executing according to the maximal coherence heuristic. The next two columns show whether a failure was detected, and whether it was diagnosed correctly. The last column shows the detection class (discussed below).

| Case | A3's Hypothesized Executing Plans | | | Relationship Failure | Diagnosis | Detection |
|---|---|---|---|---|---|---|
| # | Attacker A1 | Attacker A2 | Scout A3 | Detected? | Success? | Class |
| 1 | J | J | J | - | n/a | True Negative |
| 2 | W | J | J | + | + | True Positive |
| 3 | J | W | J | + | + | True Positive |
| 4 | H | H | J | + | - | True Positive |
| 5 | W | W | W | - | n/a | True Negative |

Table 2: Scout's (A3) monitoring results in all permutations of Example 2.

For example, case 2 in Table 2 shows the results of the scout monitoring in the original failure in Example 2 (Section 2). Using RESL, and selecting a maximally-coherent hypothesis, the scout hypothesizes that the non-moving attacker is executing W (case 2, column 2), while the moving attacker is executing J (case 2, column 3). The scout of course knows that its own selected plan J (case 2, column 4). A violation of the teamwork relationship is thus detected (case 2, column 5), since A1's W is not in agreement with the rest of the team's J. Furthermore, the diagnosis was successful in identifying the cause for the failure, i.e., the fact that the enemy's position has been identified by the scout, but no knowledge of this was passed on to the failing attacker (case 2, column 6).

The last column of Table 2 shows the detection class of each failure. The detection class of a case can be one of: true positive, true negative, false positive, and false negative. These correspond to the following possible monitoring outcomes: A true positive is an outcome where a relationship failure has actually occurred, and has been detected. A true negative is where no failure has occurred, and the system correctly reports none is being detected. A false positive is where no failure has occurred, but the system nevertheless incorrectly detects one, and a false negative is where a failure has occurred, but the system fails to detect it. Table 2 shows that in all permutations of Example 2 the teamwork monitoring techniques did not encounter the problematic false positive or false negative cases.

A closer look at these results hints at a key contribution of this paper in addressing the monitoring selectivity problem: Effective failure detection can take place despite the use of uncertain, limited, knowledge about monitored agents. In case 4 of Table 2, the monitoring agent was able to detect the failure *despite being wrong about the state of the agents involved*. The scout believes that the two attackers are executing the H (`ordered-halt`) plan, but they are actually executing, W. H is selected when a command is received from headquarters to halt execution and hover in place. From the scout's perspective, a hovering attacker can therefore be inferred to be executing H or W. Thus two equally-ranked maximally-coherent hypotheses exist: the two attackers are either both executing W or both executing H. A random selection was made, and in this case resulted in the wrong hypothesis being selected. Nevertheless, a violation of the teamwork relationships was detected, as neither H nor W agrees with the scout's J.

However, as the last column of case 4 shows (in Table 2), the diagnosis procedures are sensitive to the selection of the team-modeling hypothesis. The hypothesis used in this case does not correctly reflect the true state of the agents, and so despite the scout's success to detect a failure in this case, the diagnosis procedures fail to provide correct diagnosis (the diagnosis was successful in the two other failure cases). This phenomenon repeats in other empirical results we provide below: diagnosis failed whenever the hypothesis chosen was incorrect, although it was sufficient for detection. We intend to explore ways to improve the accuracy of the hypotheses in future work, and therefore will not address this issue in this paper anymore. The failure detection capabilities are a significant improvement in themselves, since the agents know with certainty that a failure has occurred, even if their diagnosis of it is incorrect.

Many social and physical failures are successfully captured using the team-coherence heuristic for monitoring selectivity. In fact, in our permutations of Example 2, no matter which one of the agents was the monitor, all failures to maintain relationship (i.e., all physical failures except for one in which the team remains in agreement) were detected reliably, although sometimes diagnosis failed. This result is especially surprising considering that only a single agent did the monitoring. Previous monitoring methods (condition monitors and communications) were unable to detect the failures, despite being used by all three agents .

Tables 3 and 4 present further empirical results, on the basis of Example 1. Table 3 presents all failure permutations of Example 1 in the same format as Table 1. The normal order of execution of these plans is as follows: All agents jointly execute the `fly-flight-plan` (F) plan until they detect the way-point. They then switch to the `wait-at-point` (W) plan, in which the two attackers land while the scout continues to fly ahead to identify the enemy.

The failures here are all on the part of the agents to detect the way-point, and thus to switch to the `W` plan.

| Case | Actual Executing Plans | | | Relationship Failure | Physical |
|------|------------|------------|----------|----------|----------|
| # | Attacker A1 | Attacker A2 | Scout A3 | Occurred? | Failure |
| 1 | W | W | W | - | - |
| 2 | F | W | W | + | A1 vision fails |
| 3 | W | F | W | + | A2 vision fails |
| 4 | F | F | W | + | A1, A2 vision fails |
| 5 | W | W | F | + | A3 vision fails |
| 6 | F | W | F | + | A1, A3 vision fails |
| 7 | W | F | F | + | A2, A3 vision fails |
| 8 | F | F | F | - | A1,2,3 vision fails |

Table 3: All failure permutations of the undetected way-point scenario (Example 1).

| Case | A1's Hypothesized Executing Plans | | | Relationship Failure | Detection |
|------|------------|------------|----------|----------|----------|
| # | Attacker A1 | Attacker A2 | Scout A3 | Detected? | Class |
| 1 | W | W | W | - | True Negative |
| 2 | F | W | F | + | True Positive |
| 3 | W | F | W | + | True Positive |
| **4** | **F** | **F** | **F** | - | **False Negative** |
| **5** | **W** | **W** | **W** | - | **False Negative** |
| 6 | F | W | F | + | True Positive |
| 7 | W | F | F | + | True Positive |
| 8 | F | F | F | - | True Negative |

Table 4: Attacker's (A1) monitoring results in all permutations of Example 1.

Table 4 present the monitoring results for all permutations of Example 1. Here the attacker A1 is monitoring the team using again a maximally team-coherent hypothesis in detecting failures. The results show that A1 is successful in detecting all teamwork failures but two (cases 4-5, highlighted in bold face).

These two false outcomes are both false negatives. In both these cases, the monitoring attacker A1 picked an incorrect hypothesis for the scout, since the scout's actions lead to ambiguous interpretations. The scout is to fly forward (to scout the enemy) if it detected the way-point (plan `W`), but also if it did not (then it would be flying "in formation"–plan `F`). The use of the maximal team-coherence heuristic causes A1 to prefer a hypothesis in which the scout is in agreement with the attackers when in fact it is not. For example, in case 4, the two attackers have failed to detect the way-point and are executing `F`. Observing the scout, the monitoring attacker A1 is not sure whether the scout is executing `F` or `W`. However, believing that the scout is executing `F` results in a maximally-coherent team-modeling hypothesis (all the agents are in agreement), while believing that the scout is executing `W` results in a less coherent hypothesis. Thus A1 selects a wrong hypothesis, which in this case fails to detect the teamwork failure.

The maximal team-coherence heuristic can detect failures despite using incorrect hypotheses. Unfortunately, such hypotheses can also lead to false-negatives as we have seen in Table 4. However, none of our experiments resulted in a false-positive result, i.e., a result in which the system detected a failure but in reality none had occurred. Thus the heuristic provided sound results in these cases. We are able to formally prove this property holds in general when the maximal team-coherence heuristic is used.

First, we address a matter of notation. Let an agent $A$ monitor an agent $B$, which is executing some plan $P$. We denote by $M(A, B/P)$ the set of agent-modeling hypotheses that $A$'s agent-modeling component constructs based on $B$'s observable behavior during the execution of $P$. In other words, $M(A, B/P)$ is $A$'s set of all plans that match $B$'s observable behavior. Note that when $A$ monitors itself, it has direct access to its own state and so $M(A, A/P) = \{P\}$. Using the modeling notation, we make the following definitions which ground our assumptions about the underlying knowledge used in monitoring:

**Definition 3.** Given a monitoring agent $A$, and a monitored agent $B$, we say that $A$'s *agent-modeling* of agent $B$ is *complete* if for any plan $P$ that may be executed by B, $P \in M(A, B/P)$.

The set $M(A, B/P)$ will typically include other matching hypotheses besides the correct hypothesis P, but is guaranteed to include P. Following this definition of *individual* agent-modeling completeness, we can define group-wide team-modeling completeness:

**Definition 4.** Let $A$ be an agent monitoring a team $T$ of agents $B_1, \cdots, B_n$. We say that $A$'s *team-modeling* of the team $T$ is *complete* if $A$'s agent-modeling of each of $B_1, \cdots, B_n$ is complete.

Definition 4 is critical to guarantee the capabilities we will explore analytically in this section and the next. It generally holds in our use of RESL in the ModSAF and RoboCup domains, and we make it explicit here in service of applications of the techniques in other domains.

Armed with these definitions, we now formalize the failure detection capabilities suggested by the empirical evidence in Theorem 1.

**Theorem 1.** *Let a monitoring agent $A$ monitor a simple team $T$. If $A$'s team-modeling of $T$ is complete, and $A$ uses a maximally team-coherent hypothesis for detection, then the teamwork failure detection results are sound.*

*Proof.* We will show that if no failure has occurred, none will be detected, and thus that any failure that is detected is in fact a failure. Let $a_1, \ldots, a_n$ be the agent members of $T$. Each agent $a_i$ is executing some plan $P_i$ ($1 \leq i \leq n$). Thus collectively, the group is executing $(P_1, \ldots, P_n)$. If no failure has occurred, then all the agents are executing the same plan $P_0$, i.e., $\forall i, P_i = P_0$. Since $A$'s team-modeling is complete, the correct hypothesis $(P_0, \ldots, P_0)$ is going to be in the set of team-modeling hypotheses $H$. Since it is a maximally team-coherent hypothesis, either it will be selected, or that a different hypothesis *of the same coherence level* will be selected. Any hypothesis with the same coherence level as the correct one implies no failure is detected. Thus the detection procedure is sound. $\square$

Despite uncertainty in the knowledge used, sound failure-detection can be guaranteed using the maximal team-coherence heuristic. This is one answer to the monitoring selectivity problem. However, as we have seen in Table 4, some failures may pass undetected using this heuristic (i.e., it may result in false-negatives). Detection using maximal team-coherence may therefore unfortunately be *incomplete*. We may prefer our monitoring system to be *complete*—guaranteed to detect all teamwork failures.

We therefore experimented with the maximal team-*incoherence* heuristic, the inverse of the maximal team-coherence heuristic. This heuristic prefers hypotheses that suggest *more* failures, rather than less. Table 5 gives the monitoring attacker A1's view of the team, similar to Table 4, but using a maximally team-*incoherent* hypothesis. It shows that indeed using a maximally team-incoherent hypothesis will not lead to the false-negative detections in cases 4 and 5 (in contrast to these cases in Table 4).

| Case | A1's Hypothesized Executing Plans | | | Relationship Failure | Detection |
|:---:|:---:|:---:|:---:|:---:|:---:|
| # | Attacker A1 | Attacker A2 | Scout A3 | Detected? | Class |
| 1 | **W** | **H** | **F** | + | **False Positive** |
| 2 | F | H | W | + | True Positive |
| 3 | W | F | F | + | True Positive |
| 4 | F | F | W | + | True Positive |
| 5 | W | H | F | + | True Positive |
| 6 | F | H | W | + | True Positive |
| 7 | W | F | F | + | True Positive |
| 8 | **F** | **F** | **W** | + | **False Positive** |

Table 5: Attacker's (A1) monitoring results in all permutations of Example 1, using team-*incoherence*.

Guided by these results, we formally show that the team-incoherence heuristic leads to a detection procedure that is *complete*.

**Theorem 2.** *Let a monitoring agent A monitor a simple team T. If the A's team-modeling of T is complete, and A uses a maximally team-incoherent hypothesis for detection, then the teamwork failure detection results are complete.*

*Proof.* Analogous to that of Theorem 1, the proof is provided in appendix A. □

However, these successes are offset by false positive outcomes in cases 1 and 8 of Table 5. In these cases, no failures have occurred, but the monitoring system falsely reported detected failures. In practice, this may lead to costly processing of many false alarms.

Ideally, the detection capabilities should be sound *and* complete. Unfortunately, we can show that no coherence-based disambiguation scheme exists that results in both sound and complete detection. We show in Theorem 3 that to provide sound and complete detection, a disambiguation method will have to be inconsistent: Given the same set of possible matching hypotheses, it will have to sometimes rank one hypothesis on top, and sometimes another.

**Theorem 3.** *Let H be a complete team-modeling hypotheses set, modeling a simple team. There does not exists a disambiguation scheme S that (1) uses coherence alone as the basis* *for disambiguation of H, and (2) is deterministic in its selection, and (3) results in sound and complete failure detection.*

*Proof.* Let $S$ be a disambiguation scheme that leads to complete and sound detection and uses only knowledge of the coherence of the hypotheses in selecting a disambiguated hypothesis. Suppose for contradiction that it is deterministic, and thus consistent, in its selection of an hypothesis out of $H$, i.e., given $H$, a set of candidate hypotheses, it applies some deterministic procedure to choose one hypothesis based on its coherence. Since it does not use any other knowledge outside of the coherence of the candidate hypotheses, given the same set of candidates, it will always choose the same hypothesis. Let $A_m$ be the monitoring agent using $S$. Let $B$ be a monitored agent, whose actions are identical when executing team plans $P_1, P_2$. Thus, $A_m$ cannot determine whether $B$ is executing $P_1$ or $P_2$, $M(A_m, B/P_1) = M(A_m, B/P_2) = \{P_1, P_2\}$. If $A_m$ and $B$ are both executing $P_1$, $A_m$'s hypotheses set is

$$H = \{(P_1, P_1), (P_1, P_2)\}$$

Since $S$ leads to complete and sound detection, it will choose $(P_1, P_1)$. Now, when $A_m$ and $B$ are executing $P_1$ and $P_2$, respectively, the matching hypothesis set is again $H$ as defined above. But now $S$ must select $(P_1, P_2)$. Since the same set of candidate hypothesis $H$ was used in each case, and no other information was supplied, $S$ must be non-deterministic in its selection of a disambiguated hypothesis, contradicting the assumption. $\qquad\square$

The empirical and analytical results show that our use of a single disambiguated hypothesis leads to improved, but imperfect, failure-detection results, compared to the monitoring conditions and communications previously used. The empirical results in Tables 2, 4, and 5 establish the benefits of the teamwork monitoring technique: Most physical failures were detected. However, the analytical results (Theorems 1, 2, 3) show that the results are less than perfect. The algorithms are either sound or complete, but not both. For complete monitoring, we would require additional procedures that can differentiate the true positives from the false ones, e.g., by focused communication. These procedures are often very expensive.

We can reduce the need for costly verification by letting go of our insistence on a single hypothesis, focusing instead on maintaining two hypotheses: a maximally-coherent hypothesis and a maximally-incoherent hypothesis. Table 6 shows a portion of the full set of team-hypotheses available when the attacker A1 is monitoring the team. The total number of hypotheses presented in the table is 24, with as many as 4 co-existing in a single case, and thus maintaining a full set of hypotheses would be expensive. However, the two inverse heuristics—team-coherence and incoherence—represent two extremes of the space of these hypotheses. If they agree that a failure exists, then a failure actually occurred, since the team-coherent hypothesis guarantees soundness (Theorem 1). If they agree that no failure exists, then no failure took place, since the team-incoherent hypothesis guarantees completeness (Theorem 2). If they disagree (i.e., the team-coherent hypothesis does not imply a failure, but the team-incoherent hypothesis does), the monitoring system cannot be sure either way, and must revert back to verification.

This revised detection algorithm offers significant computational savings compared to the single team-incoherent hypothesis approach. It is complete and unsound, but significantly reduces the need for verification, since at least when the team-coherent hypothesis implies failures, verification is not necessary. It requires representing only two hypotheses, and is thus still computationally cheaper than maintaining an exponential number of hypotheses.

| Case | A1's Hypothesized Executing Plans | | | Relationship Failure | Detection |
| :---: | :---: | :---: | :---: | :---: | :---: |
| # | Attacker A1 | Attacker A2 | Scout A3 | Detected? | Class |
| 1 | W | H | F | + | False Positive |
|   | W | H | W | + | False Positive |
|   | W | W | F | + | False Positive |
|   | W | W | W | - | True Negative |
| 2 | F | H | F | + | True Positive |
|   | F | H | W | + | True Positive |
|   | F | W | F | + | True Positive |
|   | F | W | W | + | True Positive |
| 3 | W | F | F | + | True Positive |
|   | W | F | W | + | True Positive |
| 4 | F | F | W | + | True Positive |
|   | F | F | F | - | False Negative |
| 5 | W | H | F | + | True Positive |
|   | W | H | W | + | True Positive |
|   | W | W | F | + | True Positive |
|   | W | W | W | - | False Negative |
| 6 | F | H | W | + | True Positive |
|   | F | H | F | + | True Positive |
|   | F | W | W | + | True Positive |
|   | F | W | F | + | True Positive |
| 7 | W | F | F | + | True Positive |
|   | W | F | W | + | True Positive |
| 8 | F | F | W | + | False Positive |
|   | F | F | F | - | True Negative |

Table 6: A portion of the attacker's (A1) monitoring hypotheses and implied results when no ranking is used to select a single hypothesis for each case.

For example, using a maximally team-incoherent hypothesis on permutations of Example 1 results in a need to verify in all eight cases as we have seen (5). However, when we combine such an hypothesis with a maximally team-coherent hypothesis (e.g., as in Table 4), we only need to verify four (50% ) of the cases. In cases 2, 3, 6, 7 there is agreement between the two hypotheses that a failure has occurred, and thus no verification is required.

A monitoring agent can therefore address the monitoring selectivity problem by balancing its resource usage against the guaranteed performance of the monitoring algorithm used. Either of the simpler single-hypothesis algorithms would utilize only one hypothesis in each case, with detection capabilities that are guaranteed to be sound or complete, but not both. In the more complex algorithm, two hypotheses would be reasoned about in each case, and the algorithm would be complete and require verification in fewer cases compared to the simple-hypothesis complete algorithm.

## 5. Monitoring Selectivity in Distributed Teamwork Monitoring

This section focuses on monitoring selectivity when exploiting a key opportunity for execution monitoring in multi-agent environments—it is not only the monitored agents that are distributed, but the monitoring agents can be distributed as well. We begin with the simple scheme of selecting a single maximally team-coherent hypothesis. Since centralized teamwork monitoring was successful in addressing all permutations of Example 2, we focus here on the permutations of Example 1 (Table 3), in which centralized teamwork monitoring by the attacker resulted in false-negative detections (cases 4-5 in Table 4).

In a distributed teamwork monitoring scheme, not only will a single attacker monitor its teammates, but the scout (and the other attacker) will also engage in monitoring. Table 7 presents the monitoring results of the same failure permutations, with the scout as the monitoring agent. We find that the scout successfully detects the two failure cases that the attacker failed to detect, compensating for the attackers' monitoring mistakes. Furthermore, since the scout used the the maximal-coherence heuristic, detection is sound and no verification is required. The reason for the scout's success is that the attackers' actions in this case, although ambiguous, do not support any hypothesis that can be matched to the scout's plan. In other words, regardless of what plan the attackers are executing in these two cases, it is different that the plan executed by the scout.

| CASE # | A3's HYPOTHESIZED EXECUTING PLANS | | | RELATIONSHIP FAILURE DETECTED? | DETECTION CLASS |
|---|---|---|---|---|---|
| | ATTACKER A1 | ATTACKER A2 | SCOUT A3 | | |
| 1 | W | W | W | - | True Negative |
| 2 | F | W | F | + | True Positive |
| 3 | W | F | W | + | True Positive |
| 4 | F | F | W | + | True Positive |
| 5 | H | H | F | + | True Positive |
| 6 | F | H | F | + | True Positive |
| 7 | H | F | F | + | True Positive |
| 8 | F | F | F | - | True Negative |

Table 7: Scout's (A3) monitoring results in all permutations of Example 1, using team-coherence.

Thus if all agents engaged in monitoring in permutations of Example 1, detection would be sound and complete. In all actual failure cases (and only in those) there would at least one team-member who detects the failure. We attempt to formally define the general conditions under which this phenomenon holds.

**Definition 5.** We say that two team-plans $P_1, P_2$, have *observably-different roles* $R_1, R_2$ if given an agent $B$ who fulfills the roles $R_1, R_2$ in the two plans, resp., any monitoring agent $A$ (different than $B$) will have $M(A, B/P_1) \cap M(A, B/P_2) = \emptyset$. We then say that $B$ has observably-different roles in $P_1$ and $P_2$, and call $B$ a *key agent*.

Intuitively, $B$ is a key agent that has observably different roles in the two plans if a monitoring agent can differentiate between $B$'s behavior in executing $P_1$ and in executing $P_2$. For instance, both attackers have observably different roles in F (in which they fly) and W (in which they land). However, they do not have observably different roles in W and H, both of which require them to land. The scout has observably different roles in W (flying) and H (landing).

The key-agent is the basis for the conditions under which a self-monitoring team will detect a failure with each agent using only team-coherence. We first prove a lemma on the conditions in which a single given agent will detect a failure. We then use this lemma to prove the conditions under which at least one agent in a given team will detect a failure.

**Lemma 1.** *Suppose a simple team $T$ is self-monitoring (all members of the team monitor each other) using the maximally team-coherent heuristic (and under the assumption that for each agent, team-modeling is complete). Let $A_1, A_2$ be monitoring agents who are members of $T$ and are executing $P_1, P_2$, respectively. $A_1$ would detect a failure in maintaining teamwork relationships with an agent $A_2$, if $A_2$ is a key-agent in $P_1, P_2$.*

*Proof.* See appendix A. ☐

$A_1$ knows that it is executing $P_1$. If $A_2$ is executing $P_2$, and is a key-agent in $P_1$ and $P_2$, then $A_1$ is guaranteed to notice that a difference exists between itself and $A_2$, since $A_2$ is acting observably different than it would if it had been executing $P_1$. Note, however, that $A_2$ may or may not detect this difference, since from $A_2$'s perspective, $A_1$'s behavior may or may not be explained by $P_2$. $A_2$ will detect a difference only if $A_1$'s roles in $P_1$ and $P_2$ are also observably-different. However, since $A_1$ has detected the failure, it can now alert its teammates, diagnose the failure, or choose corrective action.

If we want to guarantee that a teamwork failure will always be detected by at least one agent, we must make sure that in each possible combination of plans, there has to be at least one key-agent whose roles are observably different. The lemma shows that other agents monitoring this agent will notice a failure if one occurs. To this aim, we define an observably-partitioned set of plans employed by a team.

**Definition 6.** A set $P$ of team-plans is said to be *observably-partitioned* if for any two plans $P_i, P_j \in P$ there exists a key-agent $A_{ij}$. The set of these $A_{ij}$ agents is called the *key agents set of $P$*.

For instance, the set of team-plans our helicopter pilots team has been using in the examples (`Fly-Flight-Plan` (F), `Wait-at-Point` (W), `Ordered-Halt` (H), and `Join-Scout` (J)) is observably-partitioned. The attackers land in W and H, but fly in F and J. The scout lands in J and H, but flies in W and F. Table 8 shows which agents have observably different roles in any two plans in the set. For instance, by finding the cell at the intersection of the H row and the W column, we find that the scout has observably different roles in these two plans. Indeed, the scout lands when a command is received to halt execution (H), but flies out to scout the enemy's position when executing W. Here, since all agents have observably-different roles in at least two plans, the key agents set of { W, F, H, J } includes all members of the team—attackers and scout.

| | Fly-Flight-Plan (F) | Wait-at-Point (W) | Ordered-Halt (H) | Join-Scout (J) |
|---|---|---|---|---|
| F | - | Attackers | Attackers | Scout |
| W | Attackers | - | Scout | Scout and Attackers |
| H | Attackers | Scout | - | Attackers |
| J | Scout | Scout and Attackers | Attackers | - |

Table 8: Observable partitioning of the helicopter pilot team in ModSAF.

**Theorem 4.** *If a simple team (1) employs an observably-partitioned set of team-plans $O$, and all team-members monitor members of the key agents set of $O$, (2) using complete team-modeling and (3) maximally team-coherent hypotheses, then the teamwork failure detection results are sound and complete.*

*Proof.* From theorem 1 we know that detection would be sound. To show that it is complete, we will prove at least one agent will detect a difference between itself and others whenever team-members are not all executing the same plan (i.e., a failure is occurring). Suppose the team is currently divided on the team-plans that must be executed, i.e., there are agents $a_i, a_j$ in the team that are executing team plans $P_i, P_j$, respectively, such that $P_i \neq P_j$. Thus a failure has occurred. Let $K$ be the key agents set of $O$. Since the team is observably-partitioned, for $P_i, P_j$ there exists at least one key agent $a_1 \in K$. There are three cases:
**case (i).** $a_1$ is executing $P_i$. In this case any agent executing $P_j$ would detect a difference with $a_1$ and would therefore detect the failure (lemma 1).
**case (ii).** $a_1$ is executing $P_j$. In this case any agent executing $P_i$ would detect a difference with $a_1$ and would therefore detect the failure (lemma 1).
**case (iii).** $a_1$ is executing some other plan $Q$. Its roles must be observably different in $Q$ and $P_i$, or in $Q$ and $P_j$ (or in both), and thus any agent executing $P_j$ and/or $P_i$ would detect the failure. The case that $a_1$'s roles are not observably different in $Q$ and $P_i$ *and* in $Q$ and $P_j$ is impossible, since then for a monitoring agent $A_m$

$$M(A_m, a_1/P_i) \cap M(A_m, a_1/Pj) \supseteq \{Q\} \neq \emptyset$$

Contradicting $a_1$ being a key agent for $P_i, P_j$.

Since in all three cases, at least one agent would detect a failure where one occurred. Therefore, failure detection is complete. Since it is also sound as we have seen, detection is sound and complete. □

The theorem shows that distributed teamwork monitoring can result in *sound and complete* failure-detection, while using a simple algorithm. Each team-member monitors *only the key agents*[2], using a maximally team-coherent hypothesis. If it detects a failure, then certainly one has occurred. If no agent detects a failure, then indeed no failure has occurred.

This simple distributed algorithm, with its attention-focusing features and guaranteed soundness and completeness contrasts with the more complex centralized algorithm which we discussed in the previous section (Section 4). The algorithm's effectiveness relies on the condition of an observably-partitioned set of plans, and on the distribution of the monitoring. A corollary of Theorems 3 and 4 is that if key agents are not available in the distributed case, failure detection is either sound or complete, but not both. And even when key agents are available, centralized teamwork monitoring is still not complete and sound.

---

2. If the monitoring team-member does not know who the key agents are, but knows they exist, it can monitor all other team-members. This increases monitoring, but sound and complete failure detection is still guaranteed.

Fortunately, observable-partitioning is not a difficult property to design: Teams are very often composed such that not all agents have the same role in the same plan, and in general, roles do have observable differences between them. For instance, our helicopter pilot team in the ModSAF domain typically executes a set of plans with this property, as Table 8 demonstrates.

If the team, however, is not observably-partitioned, there may be a case where two agents are each executing a different plan, but no agent will be able to detect it using the team-coherence heuristic. The minimal case where this occurs is when two agents, $A_1$ and $A_2$ are executing plans $P_1$ and $P_2$, respectively, and $P_1$ and $P_2$ are not observably different, such that

$$M(A_2, A_1/P_1) \cap M(A_1, A_2/P2) = \{P_1, P_2\}$$

This will result in $A_1$ and $A_2$ each believing that the other is in agreement with them. A check for such a situation can be made a part of the plan design process, marking *risky points* in the execution in which detection is either sound or complete (Theorem 3), and verification (e.g., by communications) can be prescribed pro-actively. Or, the check could be inserted into the protocol for run-time analysis—the agent would simulate the other's hypotheses matching their own actions, and detect risky points dynamically.

## 6. Using Socially-Attentive Monitoring in an Off-Line Configuration

To further demonstrate the generality of our socially-attentive monitoring framework, this section examines re-use of teamwork monitoring in domains in which diagnosis and recovery from every failure are infeasible during execution. Examples of such domains include team sports, military human team training (Volpe, Cannon-Bowers, & Salas, 1996), and other multi-agent domains. The dynamic nature of the domain, hard real-time deadlines, and complexity of the agents involved (e.g., human team members) make diagnosis and recovery difficult. Even if a failure can be diagnosed, it is often too late for effective recovery. In such environments, the monitoring agent is often concerned with trends of performance. This information is important for long-term design evaluation and analysis, and need not necessarily be calculated on-line. The results of the analysis are meant as feedback to the agents' designer (coach or supervisor, for humans).

To this end, we are developing an off-line socially-attentive monitoring system called TEAMORE (TEAmwork MOnitoring REview). TEAMORE currently uses execution traces of the monitored agents to perform the monitoring, rather than using plan recognition. Thus it does not need to worry about the uncertainty in plan-recognition, nor about real-time performance. Instead, it knows with certainty each agent's plans during execution. TEAMORE accumulates several quantitative measures related to teamwork, including the Average-Time-to-Agreement measure (ATA, for short), and a measure of the level of agreement in a team. These build on the failure detection algorithm, but aggregate failures in quantitatively. We focus here on the ATA measure.

Teamore defines a *switch* as the time interval beginning at the point where any team-member (at least one) selects a new team plan for execution by the team, and ending at the point where the team is again in agreement on the team-plan being executed. In perfect teamwork, all team-members select a new team-plan jointly, and so always remain in agreement. In more realistic scenario, some agents will take longer to switch, and so initially a teamwork failure will occur. The first team-member to select a new plan will be in disagreement with some of its teammates, until either it rejoins them in executing the original plan, or they join it in selecting the new plan. Such a switch begins with a detected failure and ends when no more failures are detected.

Figure 6 shows an illustration of a switch. The three agents begin in an initial state of agreement on joint execution of Plan 1 (filled line). Agent 1 is the first agent to switch to Plan 2 (dotted line), and is followed by Agent 3, and finally Agent 2. The switch is the interval which begins at the instance Agents 1 selected Plan 2, to the time all three agents regained their agreement (but this time on Plan 2).

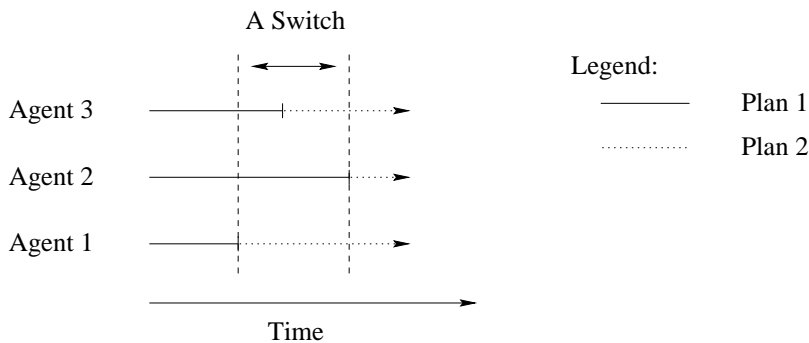

Figure 6: An illustration of a switch. The three agents switch from plan 1 to plan 2.

Teamore keeps track of the lengths of time in which failures are detected until they are resolved. The ATA measure is the average switch length (in time "ticks") per a complete team run (e.g., a mission in ModSAF, a game in RoboCup). A perfect team would have all switches of length zero, and therefore an ATA of 0. The worst team would be one that from the very beginning of their task execution to the very end of it, would not agree on the team plan being executed. For instance, each RoboCup game lasts for 6000 "ticks". The worst possible team would have only one switch during the game, of length 6000. Thus the ATA scale in RoboCup goes from 0 (perfect) to 6000 (worst).

We used the ATA measure to analyze a series of games of our two RoboCup simulation-league teams, ISIS'97 and ISIS'98 (Marsella et al., 1999) against a fixed opponent, Andhill'97 (Andou, 1998). In these games, we varied the use of communications by our teams to evaluate design decisions on the use of communications. In approximately half of the games, players were allowed to use communications in service of teamwork. In the other half, all communications between agents were disabled. ISIS'97 played approximately 15 games in each settings, and ISIS'98 played 30 games in each communication settings.

Table 9 shows the mean ATA values over these games, for two sub-teams (each having three members) of ISIS'97 and ISIS'98 (ATA values are calculated separately for each sub-team). The first column shows which sub-team the results refer to in each row. The second columns shows the mean ATA for each sub-team, when no communications were used. The third column shows the mean ATA when communications were used. The next column shows the size of the ATA reduction—the drop in the mean ATA values when communications are introduced. The last column shows the probability of the null hypothesis in a two-tailed t-test of the difference in the ATA means. This is the probability that the difference is due to chance, thus smaller numbers indicate greater significance.

| ISIS sub-team | Mean ATA No comm. | Mean ATA Comm. | ATA Reduction | t-test prob. null-hypothesis |
|---|---|---|---|---|
| '97 Goalies | 32.80 | 5.79 | 27.01 | 7.13e-13 |
| '97 Defenders | 57.5 | 6.81 | 50.69 | .45e-10 |
| '98 Goalies | 13.28 | 3.65 | 9.63 | 9.26e-16 |
| '98 Defenders | 12.99 | 3.98 | 9.01 | 7.13e-5 |

Table 9: Average-Time-to-Agreement (ATA) for games against Andhill'97.

Clearly, a very significant difference emerges between the communicating and non-communicating versions of each sub-team. The ATA values indicate that sharing information by way of communications significantly decreases the time it takes team-members to come to agreement on a selected plan. This result agrees with our intuitions about the role of communications, and in that sense, may not be surprising.

However, the ATA reduction magnitudes indicate that ISIS'98 may be much less sensitive to loss of communications than ISIS'97. The differences in ATA values for ISIS'97 are approximately triple, nearly four times, as great as for ISIS'98. Our explanation for this phenomenon is that ISIS'98 is composed of players with improved capabilities for monitoring the environment (such that they have better knowledge of the environment). ISIS'98 is therefore not as dependent on communications as are teams, such as ISIS'97, composed of players with lesser environment monitoring capabilities. ISIS'98 players are better able to select the correct plan without relying on their teammates. Thus, they would be able to maintain the same level of performance when communications are not used. In contrast, ISIS'97 players rely on passing information to and from each other (monitoring each other) through communications, and so took much longer to establish agreement when communications were not available.

We can validate the hypothesis suggested by ATA measurements by looking at the overall team-performance in the games, measured by the score difference at the end of the game. Table 10 shows the mean score difference from the same series of games against Andhill'97. The first column lists the communications settings (with or without). The second and third columns show the *mean* score-difference in the games for ISIS'97 and ISIS'98. The bottom row summarizes the results of t-tests run on each set of games, to determine the significance level of the difference between the mean score-differences. The score-difference results corroborate the ATA results. While the difference in mean score-difference is indeed statistically significant in ISIS'97 games, it is not significant in ISIS'98 games. This supports our explanation that the more situationally aware ISIS'98 is indeed better able to handle loss of communications than ISIS'97.

|  | ISIS'97 | ISIS'98 |
|---|---|---|
| Communication Used | -3.38 | -1.53 |
| Communication Not Used | -4.36 | -2.13 |
| t-test p/null hypothesis | p=0.032 | p=0.13 |

Table 10: ISIS'97 and ISIS'98 mean score difference against Andhill'97, with changing communications settings

The general lesson emerging from these experiments is that a trade-off exists in addressing the monitoring selectivity problem. The knowledge that is maintained about teammates (here, via communications) can be traded, to an extent, with knowledge maintained about the environment. A designer therefore has a range of alternative capabilities that it can choose for its agents. Different domains may better facilitate implicit coordination by monitoring the environment, while others require agents to rely on communications or explicit knowledge of team-members to handle the coordination.

The ATA results support additional conclusions, especially when combined with a general performance measure such as the score-difference. To illustrate, consider the plots of the actual data from these games. Figure 7 plots all the ATA values for all four variants, for the Goalies sub-team. The graph plots approximately 60 data-points. We see in Figure 7 that when communications are used, ISIS'97's ATA values are still generally better than ISIS'98's ATA without communications. Thus, despite its importance, individual situational awareness is not able to fully compensate for lack of communications.

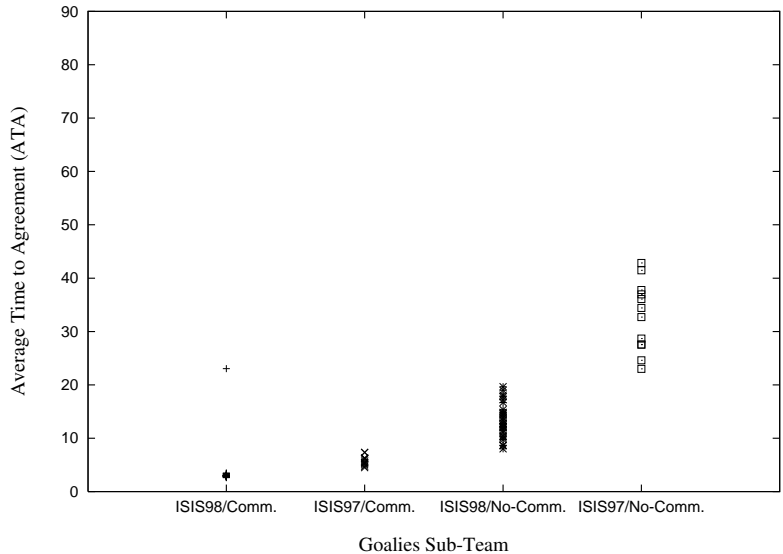

(a) ATA Values for Goalies subteam

Figure 7: ATA values for the Goalies sub-teams in games against Andhill'97.

Teamore demonstrates the reuse of the teamwork monitoring techniques developed in earlier sections in an off-line configuration. The designer of ISIS'97 should set its agents to use communications, since those will have significant improvement on the score-difference. In contrast, with or without communications, ISIS'98 players are able to maintain their collaboration. Thus if communications takes precious resources, it can be relatively safely eliminated from the ISIS'98 agents' design, and the development efforts can be directed at some other components of the agents.

## 7. Beyond Teamwork

We have presented a general socially-attentive monitoring framework to detect failures in maintaining agreement on joint team plans. However, effective operation in teams often relies on additional relationships, which we briefly address in this section.

### 7.1 A Richer Agreement Model: Agreeing to Disagree

The teamwork model requires joint execution of team plans. In service of such agreed-upon joint plans, agents may sometimes agree to execute different sub-plans individually, or split into sub-teams to execute different sub-team plans. Two examples may serve to illustrate.

**Example 3.** In the ModSAF domain, helicopters engage the enemy by repeatedly following the following three steps: hiding behind a hill or trees (*masking*), then popping up (*unmasking*), then shooting missiles at the enemy, and back to hiding. In some variations of this plan, they are required to make sure that no two helicopters are shooting at the same time. Of course, due to limits of communications, helicopters do fail and unmask at the same time.

**Example 4.** In the RoboCup domain, our 11 players in both ISIS'97 and ISIS'98 (Marsella et al., 1999) are divided into four sub-teams: mid-fielders, attackers, defenders, and goalies (the goalie and two close defenders). This division into sub-teams is modeled by the agents selecting one of four team plans in service of the `play` team plan (see Figure 8). Mid-fielders must select the `midfield` plan, goalies must select the `defend-goal` plan, etc. Again, ideally an attacker would never select any other plan but `attack`, a defender would select no other plan but `defend`, etc. However, due to communication failures, players may sometimes accidently abandon their intended sub-team, and execute a team-plan of another sub-team.

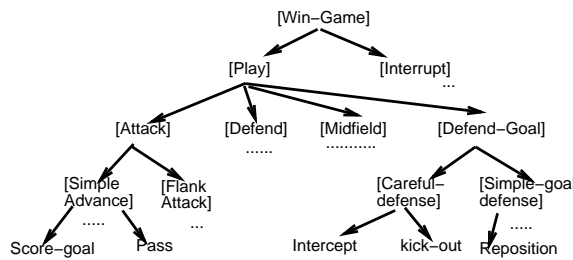

Figure 8: A Portion of the plan-hierarchy used by ISIS RoboCup agents.

In both of these examples, certain differences between agents are agreed upon and are a sign of correct execution, not of failure. Indeed, it is the lack of difference in selected plans that would indicate failure in these cases. We use the term *mutual-exclusion coordination* to refer to these relationships. In Example 3, ideally no two pilots are executing the `shooting` plan at the same time. In Example 4, no two members of different sub-teams (e.g., an attacker and a defender) are executing the same plan in service of `play` (e.g., `defend`). As the examples demonstrate, there is a clear need for monitoring mutual-exclusion coordination.

Our results of previous sections are re-used in service of socially-attentive monitoring of mutual-exclusion relationships. They require a transformation both in implementation and theory. The hierarchies are compared in the usual manner, except that failures are signified by equalities, rather than differences. For instance, if an attacker is staying in the team's own half of the field, its teammates may come to suspect that it mistakenly "defected" the attackers' sub-team and believes itself to be a defender.

The analytical results are inverted as well. The maximal team-coherence heuristic will now lead to completeness, since it prefers hypotheses that contain equalities among agents, which are failures in mutual-exclusion coordination. The maximal team-incoherence heuristic will now lead to sound detection, as it prefers hypotheses that imply no equalities have occurred. These properties can be proven formally.

**Theorem 5.** *Let a monitoring agent A monitor mutual-exclusion relationships in a group of agents G. If A's modeling of G is complete, and A uses a maximally team-incoherent hypothesis for detection, then the failure detection results are sound.*

*Proof.* Provided in appendix A. □

**Theorem 6.** *Let a monitoring agent A monitor mutual-exclusion relationships in a group of agents G. If A's modeling of G is complete, and A uses a maximally team-coherent hypothesis for detection, then the failure detection results are complete.*

*Proof.* Provided in appendix A. □

Thus in mutual-exclusion relationships, as in teamwork relationships, guaranteed failure-detection results may still be provided despite the use of limited, uncertain knowledge about monitored agents. The centralized teamwork monitoring algorithms can now be easily transformed for monitoring mutual-exclusion relationships. Unfortunately, the results in the distributed case (Theorem 4) cannot be so easily transformed, since they rely on the property of observable-partitioning, which is associated with differences, not with equalities. We leave this issue for future work.

## 7.2 Monitoring Using Role-Similarity Relationships

This section applies socially-attentive monitoring to role-similarity relationships, for monitoring individual performance within teams. In particular, in service of team-plans agents may select individual sub-plans, which do not necessitate agreement by team-members, but are constrained by the agents' roles. For instance, in service of executing the team-plan `fly-flight-plan` (Figure 3) pilots individually select their own individual plans which set the velocity and heading within the constraints of the formation and flight method specified in the mission.

Role similarity relationships specify the ways in which given individual plans are similar, and to what extent. Two agents of the same role who are executing dissimilar plans can be considered to be in violation of the role-similarity relationships. This enables a socially-attentive monitoring system to detect failure in role-execution. To monitor individual plans the agent is executing, it compares its selection with that of other agents of *the same role*, similarly to the method we used for teamwork. If the plans are considered similar by the role-similarity relationship model, no failure is detected. Otherwise, a failure may have occurred, and the diagnosis component is called to verify it and provide an explanation.

Let us illustrate with a failure from the ModSAF domain which our system was able to detect using the role-similarity relationship:

**Example 5.** A team of three helicopters was to take off from the base and head out on a mission. However, one of the pilot agents failed to correctly process the mission statement. It therefore kept its helicopter hovering above the base, while its teammates left to execute the mission by themselves.

This failures was detected using role-similarity relationship monitoring. The agreed-upon team-plan was selected by all the agents, and so no problem with teamwork relationship was detected. This team-plan involved each agent then selecting individual methods of flight, which determine altitude and velocity. Here the agents differed. The failing helicopter remained hovering, while its teammates moved forward. Using a role similarity relationship, the failing helicopter compared its own selected plan to that of its teammate (who shared its role of a subordinate in the formation), and realized that their plans were dissimilar enough to announce a possible failure.

Unfortunately, the actual similarity metrics seem to be domain- and task-specific, and thus are not as easy to re-use across domains. Furthermore, detected failures are not necessarily real failures, nor do all detected failures have the same weight. We are currently investigating ways to address these challenging issues.

## 8. Related Work

Our investigation of socially-attentive monitoring, and the relationship between knowledge maintained of agents' states and monitoring effectiveness builds on research in different sub-fields of multi-agent systems. We address these sub-fields in this section, and explain how our investigation is related to existing literature.

### 8.1 Related Work on Teamwork

Previous work in teamwork has recognized that monitoring other agents is critical to teams. Past investigations have raised the monitoring selectivity problem, but have not addressed it in depth. Building upon these investigations, this paper begins to provide some in-depth answers to this problem.

The theory of SharedPlans (Grosz & Kraus, 1996, 1999) touches on the teamwork monitoring selectivity problem in several ways, but provides only some initial answers. First, the theory requires agents to know that their teammates are capable of carrying out their tasks in the team. The authors note that agents must communicate enough about their plans to convince their teammates of their ability to carry out actions (Grosz & Kraus, 1996, p. 314). Second, the theory requires agents to have mutual-belief in the shared recipe, a state that requires agents to reason to infinite recursion about other agent's beliefs. Unfortunately, attainment of mutual belief is undecidable in theory (Halpern & Moses, 1990) and hence must be approximated in practice (Jennings, 1995; Rich & Sidner, 1997). Such approximations may still impose strong monitoring requirements. Third, theory introduces the intention-that construct in service of coordination and helpful behavior, implying monitoring of others' progress to assess the the need for such behavior (Grosz & Kraus, 1996, Axiom A5-A7). Fourth, SharedPlans requires that intentions of an agent must not conflict (Grosz & Kraus, 1996, Axiom A1), and since some of these intentions (in particular, intentions-that) may involve the attitudes of other agents, some monitoring of others to detect and avoid conflicts is implied. The authors point out that while theoretically all such conflicts can be detected, this is infeasible in practice (Grosz & Kraus, 1996, p. 307). They suggest that conflict detection and prevention be investigated in a problem-specific manner within the minimal constraints (i.e., monitoring for capabilities, mutual-belief, progress, lack of conflicts) provided by the SharedPlans framework (p. 308 and 314).

Joint-Intentions (Levesque et al., 1990; Cohen & Levesque, 1991) requires an agent who privately comes to believe that a joint-goal is either achieved, unachievable, or irrelevant, must commit to having the entire team mutually believe it to be the case. As in the theory of SharedPlans, Joint-Intentions' use of mutual belief can only be approximated in practice, and imposes strong monitoring requirements. Thus, the monitoring selectivity problem is raised for practical implementations of Joint-Intentions.

Jennings has hypothesized that two central constructs in cooperative multi-agent coordination are *commitments* made by the agents, and *conventions*, rules used to monitor these commitments (Jennings, 1993). Such conventions are used to decide what information needs to be monitored about agents, and how it is to be monitored. For instance, a convention may require an agent to report to its teammates any changes it privately detects with respect to the attainability of the team goal. Jennings raises the monitoring selectivity problem and provides an example of specific conventions for high- and low-bandwidth situations in which some knowledge is not communicated to all agents if the bandwidth is not available. However, Jennings does not explore in-depth the question of how such conventions are selected, and what are the trade-offs and guarantees associated with the selection of particular conventions. For instance, there are no guarantees on the effects of using the low-bandwidth convention in the example.

The theoretical investigations described above all raise the monitoring selectivity problem (implicitly or explicitly). Our work builds upon these to address this problem in depth, in the context of socially-attentive monitoring in teams. This paper reports on soundness and/or completeness properties of teamwork relationship failure-detection that can be analytically guaranteed, despite uncertainty in knowledge acquired about monitored agents. The analytical guarantees are applicable to plan-recognition and communications, and are corroborated by empirical results.

Building on theoretical work, practical teamwork systems include (Jennings, 1995; Rich & Sidner, 1997) and (Tambe, 1997). Jennings' investigation of the Joint-Responsibility teamwork model in GRATE* (Jennings, 1995) builds on Joint-Intentions, and similarly to our own implementation, requires agents to agree on the team-plans which are to execute. However, GRATE* is used in industrial settings in which foolproof communications can be assumed (Jennings, 1995, p. 211), and thus only passive monitoring (via communica-tions) is used. Although Jennings provides an evaluation of GRATE*'s performance with respect to communication delays, no guarantees are provided with respect to failure detection. GRATE* maintains knowledge about other agents through acquaintances models, which are used to keep track of what team-members' capabilities are (in service of forming teams). However, the question of how much knowledge should be used in these models is left unaddressed.

Rich and Sidner investigate COLLAGEN in a collaborative user-interface system, in which communications are reliable (Rich & Sidner, 1997). However, from a human-usability perspective, limiting the amount of communications is still desirable. To address this issue, recent empirical work by Lesh, Sidner and Rich (1999) utilizes plan recognition in COLLAGEN; the focus of that work is on using the collaborative settings to make the plan-recognition tractable. For instance, ambiguities in plan-recognition may be resolved by asking the user for clarification. Work on COLLAGEN does not investigate how much knowledge is to be maintained for effective collaborative dialogue with the user. In contrast, we are able to provide guarantees on the failure-detection results of our algorithms. Also, analysizing the dialogue plans for *risky points* may allow systems such as COLLAGEN to decide whether to use communications for clarification regardless of plan-recognition ambiguity.

STEAM (Tambe, 1997) maintains limited information about the ability of team-members to carry out their roles. STEAM also allows team-members to reason explicitly about the cost of communication in deciding whether to communicate or not. Our work significantly extends these capabilities via plan-recognition, and provides analytically-guaranteed fault-detection results. Furthermore, our teamwork failure-detection capabilities can be useful to trigger STEAM's re-planning capabilities.

## 8.2 Related Work on Coordination

Huber (1995) investigated the use of probabilistic plan-recognition in service of active teamwork monitoring, motivated by the unreliability and costs of passive communications-based monitoring in military applications. Washington explores observation-based coordination using Markov Models (Washington, 1998), focusing on making the computations tractable. In contrast to Huber and Washington, our work focuses on the monitoring selectivity problem. We showed strengths and limitations of centralized and distributed approaches that guaranteed failure-detection results using coherence-based disambiguation of plan-recognition hypotheses.

Durfee (1995) discusses various methods of reducing the amount of knowledge that agents need to consider in coordinating with others. The methods discussed involve pruning parts of the nested models, using communications, using hierarchies and abstractions, etc. While the focus of this work is on methods by which modeling can be limited, the focus of our work is on the question of how much modeling is required for guaranteed performance–the monitoring selectivity problem. We provide analytical guarantees on trade-offs involved in using limited knowledge of agents for failure-detection purposes.

Sugawara and Lesser (1998) report on the use of comparative reasoning/analysis techniques in service of learning and specializing coordination rules for a system in which distributed agents coordinate in diagnosing a faulty network. The investigation is focused on optimizing coordination rules to minimize inefficiency and redundancy in the agent's coordinating messages. Upon detecting sub-optimal coordination (via a fault model), the agents exchange information on their local views of the system and the problem solving activity, and construct a global view. They then compare the local view to the global view to find critical values/attributes which were missing from the local view and therefore gave rise to the sub-optimal performance problem. These values and attributes are used in constructing situation-specific rules that optimize coordination in particular situations. For example, network diagnosis agents may learn a rule that guides them to choose a coordination strategy in which only one agent performs the diagnosis and shares its result with the rest of the diagnosis agents. Our work on socially-attentive monitoring similarly uses comparison between agents views to drive the monitoring process. However, our use of comparison is a product of the relationship we are monitoring. While Sugawara and Lesser's work can be viewed as letting the agents incrementally optimize their monitoring requirements, our results analytically explore the level of monitoring required for effective failure-detection, in different configurations. Our teamwork monitoring technique addresses uncertainty in the acquired information, and does not construct a global view of all attributes the system—as that would be extremely expensive. Instead, our technique focuses on triggering failure detection via contrasting of plans, then incrementally expanding the search for differences in the diagnosis process.

Robotics literature has also raised the monitoring selectivity problem. Parker (1993) investigated the monitoring selectivity problem from a different perspective, for a formation-maintenance task. She empirically examined the effects of combining socially-attentive information (which she referred to as local) and knowledge of the team's goals, and concludes that the most fault-tolerant strategy is one where the agents monitor each other as well as progress towards the goals. Kuniyoshi et al. (Kuniyoshi, Rougeaux, Ishii, Kita, Sakane, & Kakikura, 1994) present a framework for cooperation by observations, in which robots visually attend to others as a prerequisite to coordination. The framework presents several standard attentional templates, i.e., who monitors whom. They define a team attentional structure as one in which all agents monitor each other. Our work focuses on the monitoring selectivity problem within socially-attentive monitoring of teamwork relationships, and provides analytical as well as empirical results. We treat the attentional templates as a product of the relationships that hold in the system. Our results show that monitoring in teams may not necessarily require monitoring all team-members.

## 8.3 Other Related Work

Horling et al. (Horling, Lesser, Vincent, Bazzan, & Xuan, 1999) present a distributed diagnosis system for a multi-agent intelligent home environment.The system uses fault-models to identify failures and inefficiencies in components, and to guide recovery. Schroeder and Wagner (1997) proposed a distributed diagnosis technique in which cooperating agents receive requests for tests and diagnoses, and send responses to other agents. They each construct a global diagnosis based on the local ones they produce and receive—with the assumption that no conflicts will occur. Frohlich and Nejdl (1996) investigates a scheme in which multiple diagnosis agents cooperate via a blackboard architecture in diagnosing a physical system. The agents may use different diagnosis models or systems, but a centralized conflict-resolution agent is employed to handle any conflicts in diagnoses found. All three approaches do not address the monitoring selectivity problem.

There are a few social measures related to the ATA. Goldberg and Mataric (1997) investigate a multi-robot foraging task and measure *interference*—the amount of time robots spend avoiding each other. Balch (1998) uses social entropy (Bailey, 1990) to measure *behavioral diversity* in multi-agent tasks of soccer, foraging, and formation-maintenance. Both investigations focus on characterizing heterogeneity in multi-agent systems and its relation to performance. In contrast, the focus of our work is on providing useful feedback to the designer. Possible correlation between task performance and ATA values remains to be investigated.

## 9. Conclusions and Future Work

The work presented in this paper is motivated by practical concerns. We have begun our investigation of the monitoring selectivity problem as a result of our observation that failures continue to occur despite our agents' use of monitoring conditions and communications. Analysis of the failures revealed that agents were not sufficiently informed about each other's state. While the need to monitor one's teammates has been recognized repeatedly in the past (Jennings, 1993; Grosz & Kraus, 1996; Tambe, 1997), the monitoring selectivity problem—the question of how much monitoring is required—remained largely unaddressed (Jennings, 1993; Grosz & Kraus, 1996).

We provide key answers to the monitoring selectivity problem. Within the context of socially-attentive monitoring in teams, we demonstrate that teamwork relationship failures can be detected effectively even with uncertain, limited, knowledge of team-members' states. We show analytically that centralized active teamwork monitoring provides failure-detection that is either complete and unsound, or sound and incomplete. However, centralized teamwork monitoring requires multiple hypotheses and monitoring of all team-members. In contrast, distributed active teamwork monitoring results in complete and sound failure-detection, despite using a simpler algorithm and monitoring only key agents in a team.

Using an implemented general framework for socially-attentive monitoring, we empirically validate these results in the ModSAF domain. We also provide initial results in monitoring mutual-exclusion and role-similarity relationships, and initial diagnosis procedures. We further demonstrate the generality of the framework by applying it in the RoboCup domain, in which we show how useful quantitative analysis can be generated off-line. Both ModSAF and RoboCup are dynamic, complex, multi-agent domains that involve many uncertainties in perception and action.

We attempted to demonstrate how the results and techniques can be applied in other domains. We have explicitly pointed out necessary conditions for the theorems to hold, such as observable-partitioning and team-modeling completeness. The presented diagnosis algorithm is sensitive to the accuracy of the knowledge used, and may require assuming that plans can be recognized as soon as they are selected. These conditions should be verified by the designer in the target application domain. Reactive plans (our chosen representation) are commonly used in many dynamic multi-agent domains. Our focus on monitoring agreements on joint plans stems from the centrality of similar notions of agreement in agent and human teamwork literature (Jennings, 1995; Grosz & Kraus, 1996; Volpe et al., 1996; Tambe, 1997).

We made several references to additional areas in which we would like to conduct further investigations. One important topic which we plan to investigate in depth is the strong requirements of the distributed teamwork monitoring algorithm in terms of observability. In order to provide its soundness and completeness guarantees, the distributed algorithm relies on the ability of all team-members to monitor the key agents. We are investigating ways to relax this requirement while still providing guaranteed results. In addition, the diagnosis procedures should be extended and formalized, and we would like to investigate ways to alleviate the sensitivity of these procedures to the choice of team-modeling hypothesis.

## Acknowledgments

This article is partially based on an AAAI-98 paper (Kaminka & Tambe, 1998), and an Agents-99 paper (Kaminka & Tambe, 1999) by the same authors. This research was supported in part by NSF Grant ISI-9711665, and in part by AFOSR contract #F49620-97-1-0501. We thank Jeff Rickel, George Bekey, Victor Lesser, Dan O'Leary, and David Pynadath for many useful comments. The anonymous reviewers have our thanks for helping us to crystallize our ideas and contributions in revisions of this paper.

## Appendix A. Proofs

**Theorem.** (# 2, page 123). *Let a monitoring agent $A$ monitor a simple team $T$. If $A$'s team-modeling of $T$ is complete, and $A$ uses a maximally team-coherent hypothesis for detection, then the teamwork failure detection results are sound.*

*Proof.* We will show that any failure that occurs is detected, and thus that all failures will be detected. Let $a_1, \ldots, a_n$ be the agent members of $T$. Each agent $a_i$ is executing some plan $P_i$ ($1 \leq i \leq n$). Thus collectively, the group is executing $(P_1, \ldots, P_n)$. If a failure has occurred, then there are two agents $a_k, a_j, 1 \leq j, k \leq n$ such that $a_j$ is executing plan $P_j$ and $a_k$ is executing plan $P_k$ and $P_j \neq P_k$. Since $A$'s team-modeling is complete, the correct hypothesis $(P_1, \ldots, P_j, \ldots, P_k, \ldots P_n)$ will in the set of team-modeling hypotheses. Since $A$ will choose a maximally team-incoherent hypothesis, either it will choose the correct hypothesis, which is more incoherent than a hypothesis implying no failure has occurred, or that it will select a hypothesis with greater incoherence hypothesis (or equivalent level). In any case, a failure would be detected, and the detection procedure is complete. $\square$

**Lemma.** (# 1, page 127). *Suppose a simple team $T$ is self-monitoring (all members of the team monitor each other) using the maximally team-coherent heuristic (and under the assumption that for each agent, team-modeling is complete). A monitoring agent $A_1$ who is a member of $T$ and is executing $P_1$ would detect a failure in maintaining teamwork relationships with an agent $A_2$ (also a member of $T$) executing a different plan $P_2$, if $A_2$ has an observably different role in $P_1$ and $P_2$.*

*Proof.* $A_1$ knows that it is executing $P_1$. Since all members of $T$ monitor each other and themselves, $A_1$ is monitoring $A_2$, who has an observably different role in $P_1$ and $P_2$. Since $A_2$ is executing $P_2$, and following the observably different role, $P_1 \notin M(A_1, A_2/P_2)$. Therefore from the perspective of $A_1$, it cannot be the case that it assigns $P_1$ in any *agent-modeling* hypothesis, and therefore any *team-modeling* hypothesis that $A_1$ has will have $A_1$ executing $P_1$, and $A_2$ executing some plan other than $P_1$. In other words, from $A_1$'s perspective there is no team-coherent hypothesis, and so a difference would be detected between $A_1$ and $A_2$. □

**Theorem.** *(# 5, page 134). Let a monitoring agent $A$ monitor mutual-exclusion relationships in a group of agents $G$. If $A$'s modeling of $G$ is complete, and $A$ uses a maximally team-in*coherent *hypothesis for detection, then the failure detection results are sound.*

*Proof.* We will show that if no failure has occurred, none will be detected, and thus that any failure that is detected is in fact a failure. Let $a_1, \ldots, a_n$ be the agent members of $G$. Each agent $a_i$ is executing some plan $P_i$ $(1 \leq i \leq n)$. Thus collectively, the group is executing $(P_1, \ldots, P_n)$. If no failure has occurred, then each agent is executing a different plan $(i \neq j \Rightarrow P_i \neq P_j)$. Since $A$'s group-modeling is complete, the correct hypothesis is going to be in the set of group-modeling hypotheses $H$. Since it is a maximally incoherent hypothesis, either it will be selected, or that a different hypothesis *of the same coherence level* will be selected. Any hypothesis with the same coherence level as the correct one implies no failure is detected. Thus the detection procedure is sound. □

**Theorem.** *(# 6, page 134). Let a monitoring agent $A$ monitor mutual-exclusion relationships in a group of agents $G$. If $A$'s modeling of $G$ is complete, and $A$ uses a maximally team-coherent hypothesis for detection, then the failure detection results are complete.*

*Proof.* We will show that any failure that occurs is detected, and thus that the procedure is complete. Let $a_1, \ldots, a_n$ be the agent members of $G$. Each agent $a_i$ is executing some plan $P_i$ $(1 \leq i \leq n)$. Thus collectively, the group is executing $(P_1, \ldots, P_n)$. If a failure has occurred, then there are two agents $a_k, a_j, 1 \leq j, k \leq n$ such that $a_j$ is executing plan $P_j$ and $a_k$ is executing plan $P_k$ and $P_j = P_k$. Since $A$'s group-modeling is complete, the correct hypothesis $(P_1, \ldots, P_j, \ldots, P_k, \ldots P_n)$ will in the set of group-modeling hypotheses. Since $A$ will choose a maximally team-coherent hypothesis, either it will choose the correct hypothesis, which is more coherent than a hypothesis implying no failure has occurred, or that it will select a hypothesis with greater coherence hypothesis (or equivalent level). In any case, a failure would be detected. Therefore, the detection procedure is complete. □

## Appendix B. Socially-Attentive Monitoring Algorithms

We bring here the algorithms (in pseudo-code) for the RESL plan-recognition algorithm, the comparison test supporting detection in both simple and non-simple teams, and the monitoring algorithms for the centralized and distributed cases.

### B.1 RESL

RESL works by first expanding the complete operator hierarchy for the agents being modeled, tagging all plans as non-matching. All plans' preconditions and termination conditions are flagged as non-matching as well. All plans' actions are set to be used as expectations on behavior. After initializing the plan-recognition hierarchy for each monitored agent, observations of an agent are continuously matched against the actions expected by the plans. Plans whose expectations match observations are tagged as matching, and these flags are propagated along the hierarchy, up and down, so that complete paths through the hierarchy are flagged as matching or not. These paths specify the possible matching interpretations of the observations. In addition, precondition and termination conditions are flagged as true or not, signifying the inferred appropriate belief by the modeled agents. This process is described in Algorithm 1.

---

**Algorithm 1** RESL's main loop, matching observation and making inferences for a given plan-recognition hierarchy (a single agent).

1. Get observations about agent

2. For each plan that has a set of expected observations:

   (a) Compare observations to expectations
   (b) If succeed, flag plan as matching successfully, otherwise flag plan as failing to match

3. For each plan that is flagged as matching successfully

   (a) Flag its parents as matching successfully // propagate matching

4. For each plan whose children (all of them) are flagged as failing to match

   (a) Flag it as failing to match // propagate non-matching

---

## B.2 Detection of Failure, Centralized and Distributed Teamwork Monitoring

Algorithm 2 shows how comparison of hierarchical plans is carried out. We limit ourselves here to simple-teams. The algorithm accepts as input two sets of hierarchical plan hypotheses, and their two associated agents (for clarity, the algorithms assume only two agents. The generalization to $n$ agents is straightforward). The algorithm also accepts a policy flag, `Policy`. An `OPTIMISTIC` policy causes the algorithm to use maximal team-coherence to provide sound, but incomplete detection. A `PESSIMISTIC` policy causes the algorithm to use maximal team-incoherence to provide complete, but unsound detection.

The set of hierarchical plans are marked `hierarchy_1` and `hierarchy_2`. The two agents are marked `agent_1` and `agent_2`. The algorithm makes use of the predicate `Sub-team`, which is true if the two agents (`Agent1, Agent2`) belong to different sub-teams at the given level of the hierarchy (`Depth`).

With the aid of Algorithm 2, we can now define the centralized and distributed failure detection algorithms. The centralized teamwork monitoring algorithm (Algorithm 3) utilizes Algorithm 2 twice, checking for failures with both `PESSIMISTIC` and `OPTIMISTIC` policies. If the results of both policies agree, they are certain. If the results do not agree, (i.e., the `PESSIMISTIC` policy causes a failure to be detected, while the `OPTIMISTIC` policy causes no failure to be detected), then the monitoring agent cannot be certain that a failure has taken place, and therefore needs to verify the failure. Algorithm 3 therefore returns `FAILURE, NO_FAILURE, POSSIBLE_FAILURE`.

The distributed monitoring algorithm is not given in pseudo-code form, because it is nothing more than a call to Algorithm 2 with an `OPTIMISTIC` policy parameter. Its power

---

**Algorithm 2** Hierarchical comparison of two agents, allowing for sub-teams.

1. Set Depth to 0 // look for top-most difference first

2. While both plans at depth Depth are team-plans Do

   (a) if Policy == OPTIMISTIC

       i. then Let Plan_1, Plan_2 be maximally team-coherent plans at level Depth of hierarchy_1 and hierarchy_2, respectively.

       ii. else Let Plan_1, Plan_2 be maximally team-*in*coherent plans at level Depth of hierarchy_1 and hierarchy_2, respectively.

   (b) If Plan_1 is *not equal* to Plan_2

       i. then return FAILURE

       ii. else if bottom of hierarchies reached, return NO_FAILURE, otherwise increase Depth and go to 2.

3. If only one plan is a team-plan, return FAILURE, else return NO_FAILURE.

---

**Algorithm 3** Centralized Teamwork Monitoring, applying both optimistic and pessimistic views.

1. Let Optimistic_Result = Detect(agent_1, agent_2, hierarchies_1, hierarchies_2, OPTIMISTIC)
   /* algorithm 2 */

2. Let Pessimistic_Result = Detect(agent_1, agent_2, hierarchies_1, hierarchies_2, PESSIMISTIC)
   /* algorithm 2 */

3. if Optimistic_Result == Pessimistic_Result

4. then return Optimistic_Result /* either FAILURE, or NO_FAILURE */

5. else return POSSIBLE_FAILURE

---

is derived from the fact that all members of the team are using it to monitor the key agents of the team.

# References

Ambros-Ingerson, J. A., & Steel, S. (1988). Intergrating planning, execution and monitoring. In *Proceedings of the Seventh National Conference on Artificial Intelligence (AAAI-88)* Minneapolis/St. Paul, MN. AAAI Press.

Andou, T. (1998). Refinement of soccer agents' positions using reinforcement learning. In Kitano, H. (Ed.), *RoboCup-97: Robot soccer world cup 1*, Vol. LNAI 1395, pp. 373–388. Springer-verlag.

Atkins, E. M., Durfee, E. H., & Shin, K. G. (1997). Detecting and reacting to unplanned-for world states. In *Proceedings of the Fourteenth National Conference on Artificial Intelligence (AAAI-97)*, pp. 571–576 Providence, RI. AAAI Press.

Bailey, K. D. (1990). *Social Entropy Theory*. State University of New York Press.

Balch, T. (1998). *Behavioral Diversity in Learning Robot Teams*. Ph.D. thesis, Georgia Institute of Technology.

Calder, R. B., Smith, J. E., Courtemanche, A. J., Mar, J. M. F., & Ceranowicz, A. Z. (1993). Modsaf behavior simulation and control. In *Proceedings of the Third Conference on Computer Generated Forces and Behavioral Reresentation* Orlando, Florida. Institute for Simulation and Training, University of Central Florida.

Cohen, P. R., Amant, R. S., & Hart, D. M. (1992). Early warnings of plan failure, false positives, and envelopes: Experiments and a model. Tech. rep. CMPSCI Technical Report 92-20, University of Massachusetts.

Cohen, P. R., & Levesque, H. J. (1991). Teamwork. *Nous, 35*.

Doyle, R. J., Atkinson, D. J., & Doshi, R. S. (1986). Generating perception requests and expectations to verify the execution of plans. In *Proceedings of the Fifth National Conference on Artificial Intelligence (AAAI-86)*.

Durfee, E. H. (1995). Blissful ignorance: Knowing just enough to coordinate well. In *Proceedings of the First International Conference on Multiagent Systems (ICMAS-95)*, pp. 406–413.

Fenster, M., Kraus, S., & Rosenschein, J. S. (1995). Coordination without communication: Experimental validation of focal point techniques. In *Proceedings of the First International Conference on Multiagent Systems (ICMAS-95)*, pp. 102–108 California, USA.

Firby, R. J. (1987). An investigation into reactive planning in complex domains. In *Proceedings of the Sixth National Conference on Artificial Intelligence (AAAI-87)*.

Frohlich, P., & Nejdl, W. (1996). Resolving conflicts in distributed diagnosis. In Wahlster, W. (Ed.), *the 12th Europeach Conference on Artificial Intelligence (ECAI-96)*. John Wiley & Sons, Inc.

Goldberg, D., & Mataric, M. J. (1997). Interference as a tool for designing and evaluating multi-robot controllers. In *Proceedings of the Fourteenth National Conference on Artificial Intelligence (AAAI-97)*, pp. 637–642 Providence, RI. AAAI Press.

Grosz, B. J., & Kraus, S. (1999). The evolution of sharedplans. In Wooldridge, M., & Rao, A. (Eds.), *Foundations and Theories of Rational Agency*, pp. 227–262.

Grosz, B. J., & Kraus, S. (1996). Collaborative plans for complex group actions. *Artificial Intelligence*, *86*, 269–358.

Grosz, B. J., & Sidner, C. L. (1990). Plans for discourse. In Cohen, P. R., Morgan, J., & Pollack, M. (Eds.), *Intentions in Communication*, pp. 417–445. MIT Press, Cambridge, MA.

Halpern, J. Y., & Moses, Y. (1990). Knowledge and common knowledge in a distributed environment. *distributed computing*, *37*(3), 549–587.

Hamscher, W., Console, L., & de Kleer, J. (Eds.). (1992). *Readings in Model-Based Diagnosis*. Morgan Kaufmann Publishers, San Mateo, CA.

Horling, B., Lesser, V. R., Vincent, R., Bazzan, A., & Xuan, P. (1999). Diagnosis as an integral part of multi-agent adaptability. Tech. rep. CMPSCI Technical Report 1999-03, University of Massachusetts/Amherst.

Huber, M. J., & Durfee, E. H. (1995). On acting together: Without communication. In *Working Notes of the AAAI Spring Symposium on Representing Mental States and Mechanisms*, pp. 60–71 Stanford, CA.

Jennings, N. R. (1993). Commitments and conventions: the foundations of coordination in multi-agent systems. *Knowledge Engineering Review*, *8*(3), 223–250.

Jennings, N. R. (1995). Controlling cooperative problem solving in industrial multi-agent systems using joint intentions. *Artificial Intelligence*, *75*(2), 195–240.

Johnson, W. L., & Rickel, J. (1997). STEVE: An animated pedagogical agent for procedural training in virtual environments. *SIGART Bulletin*, *8*(1-4), 16–21.

Kaminka, G. A., & Tambe, M. (1998). What's wrong with us? Improving robustness through social diagnosis. In *Proceedings of the Fifteenth National Conference on Artificial Intelligence (AAAI-98)*, pp. 97–104 Madison, WI. AAAI Press.

Kaminka, G. A., & Tambe, M. (1999). I'm OK, You're OK, We're OK: Experiments in distributed and centralized social monitoring and diagnosis. In *Proceedings of the Third International Conference on Autonomous Agents (Agents-99)* Seattle, WA. ACM Press.

Kinny, D., Ljungberg, M., Rao, A., Sonenberg, E., Tidhar, G., & Werner, E. (1992). Planned team activity. In Castelfranchi, C., & Werner, E. (Eds.), *Artificial Social Systems, Lecture notes in AI 830*, pp. 227–256. Springer Verlag, New York.

Kitano, H., Tambe, M., Stone, P., Veloso, M., Coradeschi, S., Osawa, E., Matsubara, H., Noda, I., & Asada, M. (1997). The RoboCup synthetic agent challenge '97. In *Proceedings of the International Joint Conference on Artificial Intelligence (IJCAI-97)* Nagoya, Japan.

Kraus, S., Sycara, K., & Evenchik, A. (1998). Reacing agreements through negotiations: a logical model and implementation. *artificial intelligence, 104*(1-2), 1–69.

Kuniyoshi, Y., Rougeaux, S., Ishii, M., Kita, N., Sakane, S., & Kakikura, M. (1994). Co-operation by observation – the framework and the basic task patterns. In *the IEEE International Conference on Robotics and Automation*, pp. 767–773 San-Diego, CA. IEEE Computer Society Press.

Lesh, N., Rich, C., & Sidner, C. L. (1999). Using plan recognition in human-computer collaboration. In *Proceedings of the Seventh International Conference on User Modelling (UM-99)* Banff, Canada.

Levesque, H. J., Cohen, P. R., & Nunes, J. H. T. (1990). On acting together. In *Proceedings of the Eigth National Conference on Artificial Intelligence (AAAI-90)* Menlo-Park, CA. AAAI Press.

Malone, T. W., & Crowston, K. (1991). Toward an interdisciplinary theory of coordination. Tech. rep. CCS TR#120 SS WP# 3294-91-MSA, Massachusetts Institute of Technology.

Marsella, S. C., Adibi, J., Al-Onaizan, Y., Kaminka, G. A., Muslea, I., Tallis, M., & Tambe, M. (1999). On being a teammate: Experiences acquired in the design of robocup teams.. In *Proceedings of the Third International Conference on Autonomous Agents (Agents-99)* Seattle, WA. ACM Press.

Newell, A. (1990). *Unified Theories of Cognition*. Harvard University Press, Cambridge, Massachusetts.

Parker, L. E. (1993). Designing control laws for cooperative agent teams. In *the Proceedings of the IEEE Robotics and Automation Conference*, pp. 582–587 Atlanta, GA.

Rao, A. S. (1994). Means-end plan recognition – towards a theory of reactive recognition. In *Proceedings of the International Conference on Knowledge Representation and Reasoning (KR-94)*, pp. 497–508.

Reece, G. A., & Tate, A. (1994). Synthesizing protection monitors from causal structure. In *Proceedings of Artificial Intelligence Planning Systems (AIPS-94)* Chicago, IL.

Rich, C., & Sidner, C. L. (1997). COLLAGEN: When agents collaborate with people. In Johnson, W. L. (Ed.), *Proceedings of the First International Conference on Autonomous Agents (Agents-97)*, pp. 284–291 Marina del Rey, CA. ACM Press.

Schroeder, M., & Wagner, G. (1997). Distributed diagnosis by vivid agents. In *Proceedings of the First International Conference on Autonomous Agents (Agents-97)*, pp. 268–275 Marina del Rey, CA. ACM Press.

Sugawara, T., & Lesser, V. R. (1998). Learning to improve coordinated actions in cooperative distributed problem-solving environments. *Machine Learning*, *33*(2/3), 129–153.

Tambe, M. (1996). Tracking dynamic team activity. In *Proceedings of the National Conference on Artificial Intelligence (AAAI)*.

Tambe, M. (1997). Towards flexible teamwork. *Journal of Artificial Intelligence Research*, *7*, 83–124.

Tambe, M., Johnson, W. L., Jones, R., Koss, F., Laird, J. E., Rosenbloom, P. S., & Schwamb, K. (1995). Intelligent agents for interactive simulation environments. *AI Magazine*, *16*(1).

Toyama, K., & Hager, G. D. (1997). If at first you don't succeed.... In *Proceedings of the Fourteenth National Conference on Artificial Intelligence (AAAI-97)*, pp. 3–9 Providence, RI.

Veloso, M., Pollack, M. E., & Cox, M. T. (1998). Rationale-based monitoring for planning in dynamic environments. In *Proceedings of Artificial Intelligence Planning Systems (AIPS-98)* Pittsburgh, PA.

Volpe, C. E., Cannon-Bowers, J. A., & Salas, E. (1996). The impact of cross-training on team functioning: An empirical investigation. *human factors*, *38*(1), 87–100.

Washington, R. (1998). Markov tracking for agent coordination. In *Proceedings of the Second International Conference on Autonomous Agents (Agents-98)*, pp. 70–77 Minneapolis/St. Paul, MN. ACM Press.